\documentclass[twocolumn,tighten,usenames,dvipsnames]{aastex63}
\usepackage[normalem]{ulem}
\usepackage{verbatim}
\usepackage{amssymb}
\usepackage{amsmath}
\usepackage{xspace}
\usepackage{graphicx}

\usepackage{amsmath,mathtools,mathrsfs}
\usepackage{hyperref}
\usepackage{afterpage}
\usepackage{booktabs}

\usepackage[encapsulated]{CJK}
\usepackage{ucs}
\usepackage[utf8x]{inputenc}

\def\muas{\mu{\rm as}} 
\newcommand{\themis}{{\sc Themis}\xspace}

\newcommand{\sgra}{Sgr~A$^{\ast}$\xspace}
\newcommand{\C}{\texttt{C}\xspace}

\renewcommand{\L}{{\mathcal L}}

\newcommand{\VirA}{M87*\xspace}
\newcommand{\SgrA}{Sgr~A*\xspace}

\defcitealias{M87_PaperI}{Paper~I}
\defcitealias{M87_PaperII}{Paper~II}
\defcitealias{M87_PaperIII}{Paper~III}
\defcitealias{M87_PaperIV}{Paper~IV}
\defcitealias{M87_PaperV}{Paper~V}
\defcitealias{M87_PaperVI}{Paper~VI}

\shorttitle{Measuring Spin from Relative Photon Ring Sizes}
\shortauthors{Broderick et al.}

\begin{document}
  
\title{Measuring Spin from Relative Photon Ring Sizes}

\correspondingauthor{Avery E. Broderick}
\email{abroderick@perimeterinstitute.ca}

\author[0000-0002-3351-760X]{Avery E. Broderick}
\affiliation{ Perimeter Institute for Theoretical Physics, 31 Caroline Street North, Waterloo, ON, N2L 2Y5, Canada}
\affiliation{ Department of Physics and Astronomy, University of Waterloo, 200 University Avenue West, Waterloo, ON, N2L 3G1, Canada}
\affiliation{ Waterloo Centre for Astrophysics, University of Waterloo, Waterloo, ON N2L 3G1 Canada}

\author[0000-0003-3826-5648]{Paul Tiede}
\affiliation{ Perimeter Institute for Theoretical Physics, 31 Caroline Street North, Waterloo, ON, N2L 2Y5, Canada}
\affiliation{ Department of Physics and Astronomy, University of Waterloo, 200 University Avenue West, Waterloo, ON, N2L 3G1, Canada}
\affiliation{ Waterloo Centre for Astrophysics, University of Waterloo, Waterloo, ON N2L 3G1 Canada}

\author[0000-0002-5278-9221]{Dominic W. Pesce}
\affiliation{Center for Astrophysics $|$ Harvard \& Smithsonian, 60 Garden Street, Cambridge, MA 02138, USA}
\affiliation{ Black Hole Initiative at Harvard University, 20 Garden Street, Cambridge, MA 02138, USA}

\author[0000-0003-2492-1966]{Roman Gold}
\affiliation{CP3 origins $|$ Southern Denmark University (SDU)
Campusvej 55, Odense, Denmark}

\begin{abstract}
  The direct detection of a bright, ring-like structure in horizon-resolving images of \VirA by the Event Horizon Telescope is a striking validation of general relativity. The angular size and shape of the ring is a degenerate measure of the location of the emission region, mass, and spin of the black hole.  However, we show that the observation of multiple rings, corresponding to the low-order photon rings, can break this degeneracy and produce mass and spin measurements independent of the shape of the rings.  We describe two potential experiments that would measure the spin.  In the first, observations of the direct emission and $n=1$ photon ring are made at multiple epochs with different emission locations.  This method is conceptually similar to spacetime constraints that arise from variable structures (or hot spots) in that it breaks the near-perfect degeneracy between emission location, mass, and spin for polar observers using temporal variability.  In the second, observations of the direct emission, $n=1$ and $n=2$ photon rings are made during a single epoch.
  For both schemes, additional observations comprise a test of general relativity.  Thus, comparisons of Event Horizon Telescope observations in 2017 and 2018 may be capable of producing the first horizon-scale spin estimates of \VirA inferred from strong lensing alone.  Additional observation campaigns from future high-frequency, Earth-sized and space-based radio interferometers can produce high-precision tests of general relativity.
\end{abstract}


\section{Introduction} \label{sec:intro}
The images of \VirA on horizon scales closely match the theoretical expectations from accreting supermassive black holes, providing the best evidence thus far for the identification of black holes with active galactic nuclei \citep[][hereafter Papers I-VI]{M87_PaperI,M87_PaperII,M87_PaperIII,M87_PaperIV,M87_PaperV,M87_PaperVI}.  The size of the annular emission region provided a mass estimate for \VirA of $(6.5\pm0.7)\times10^9 M_\odot$ from the dynamics of radio waves on scales of the photon orbit, the near-horizon spherical region associated with photons that execute unstable closed orbits \citepalias{M87_PaperVI}.  This mass is consistent with that implied by stellar motions at $\gtrsim 50$~pc scales, $(6.6\pm0.4)\times10^9 M_\odot$ \citep{Gebhardt2011}, confirming that gravity operates as predicted by general relativity about supermassive black holes on scales ranging from the event horizon to interstellar distances, covering more than five orders of magnitude.

\begin{figure*}
\begin{center}
\includegraphics[width=0.32\textwidth]{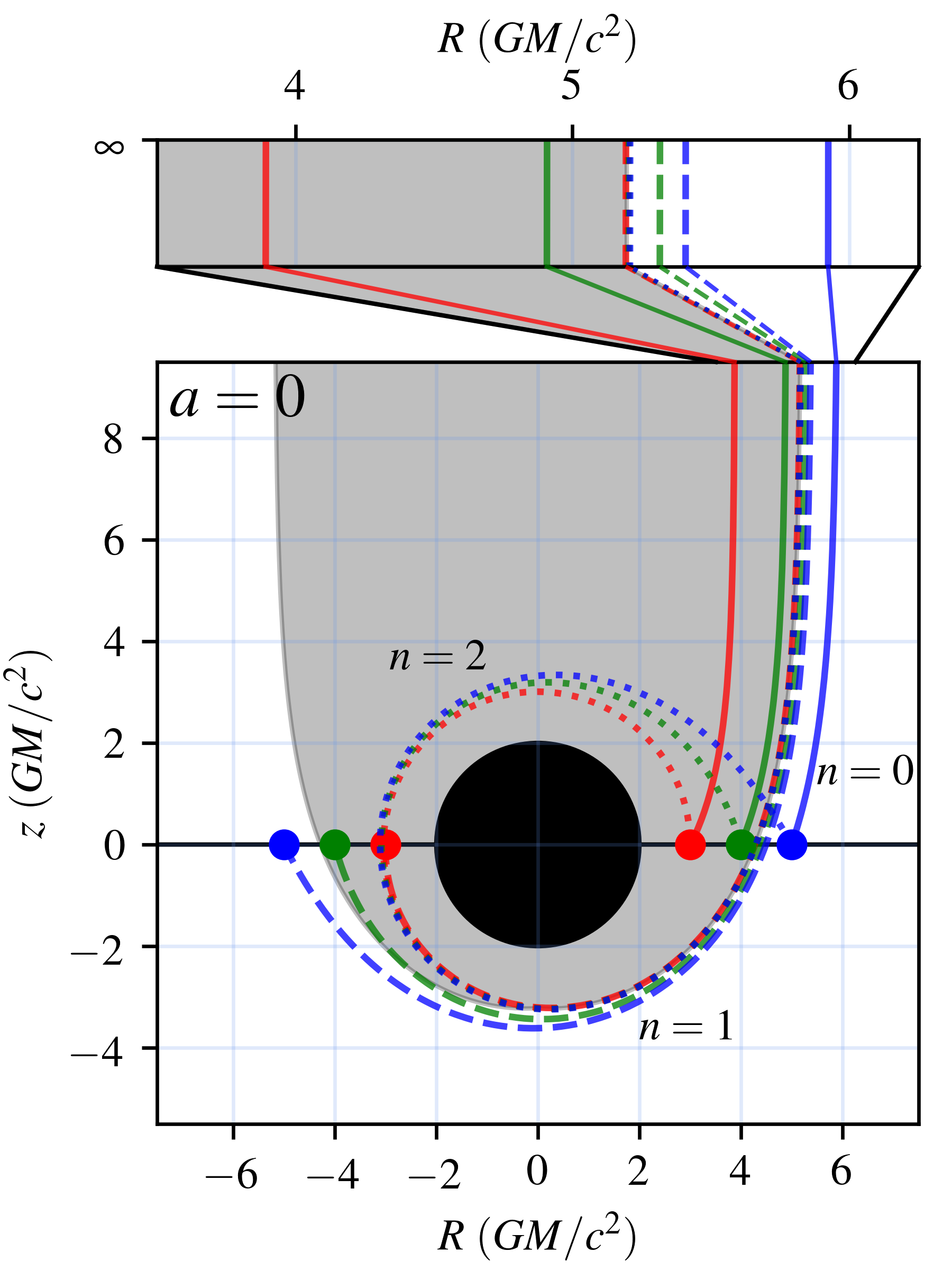}
\includegraphics[width=0.32\textwidth]{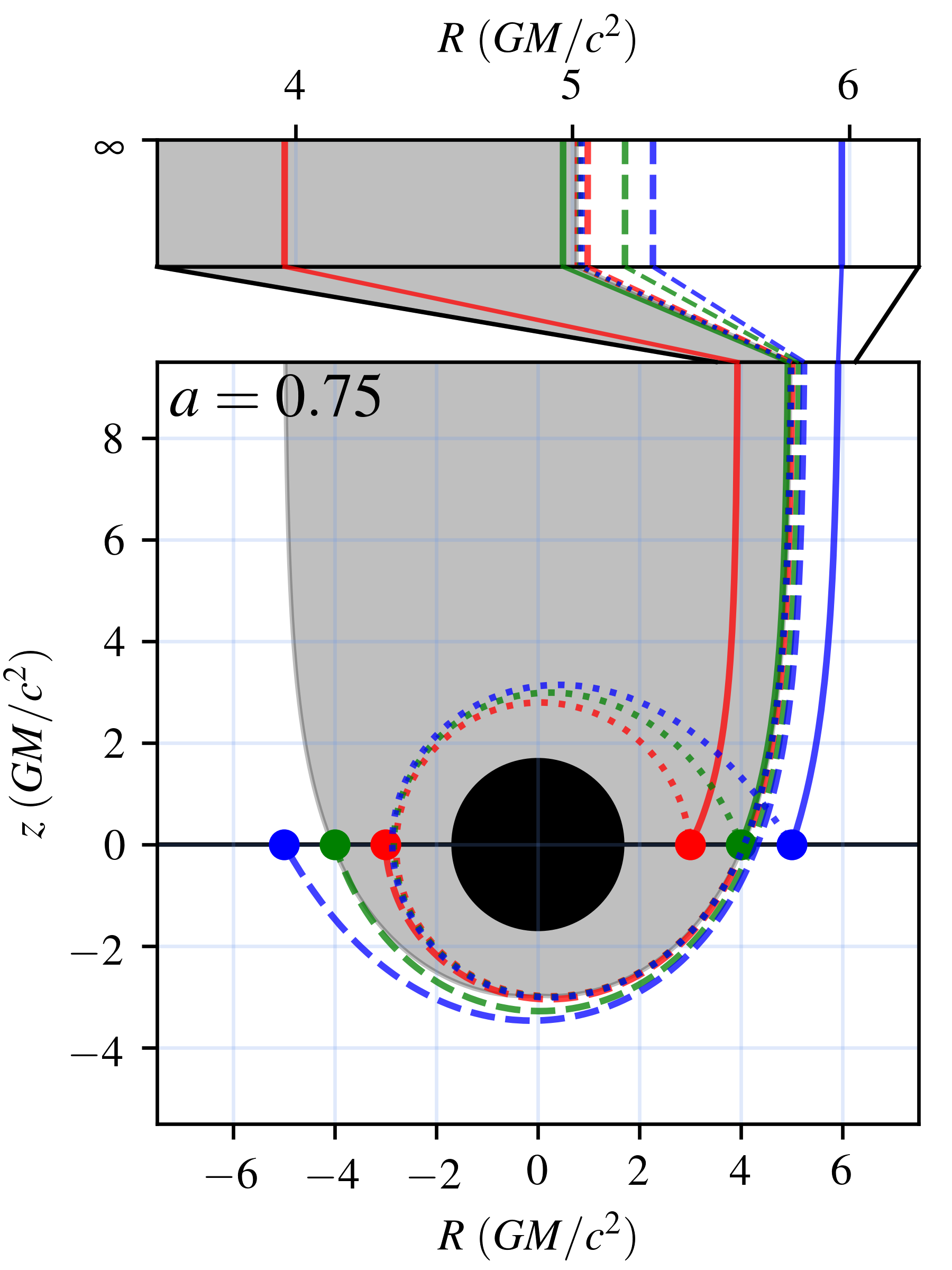}
\includegraphics[width=0.32\textwidth]{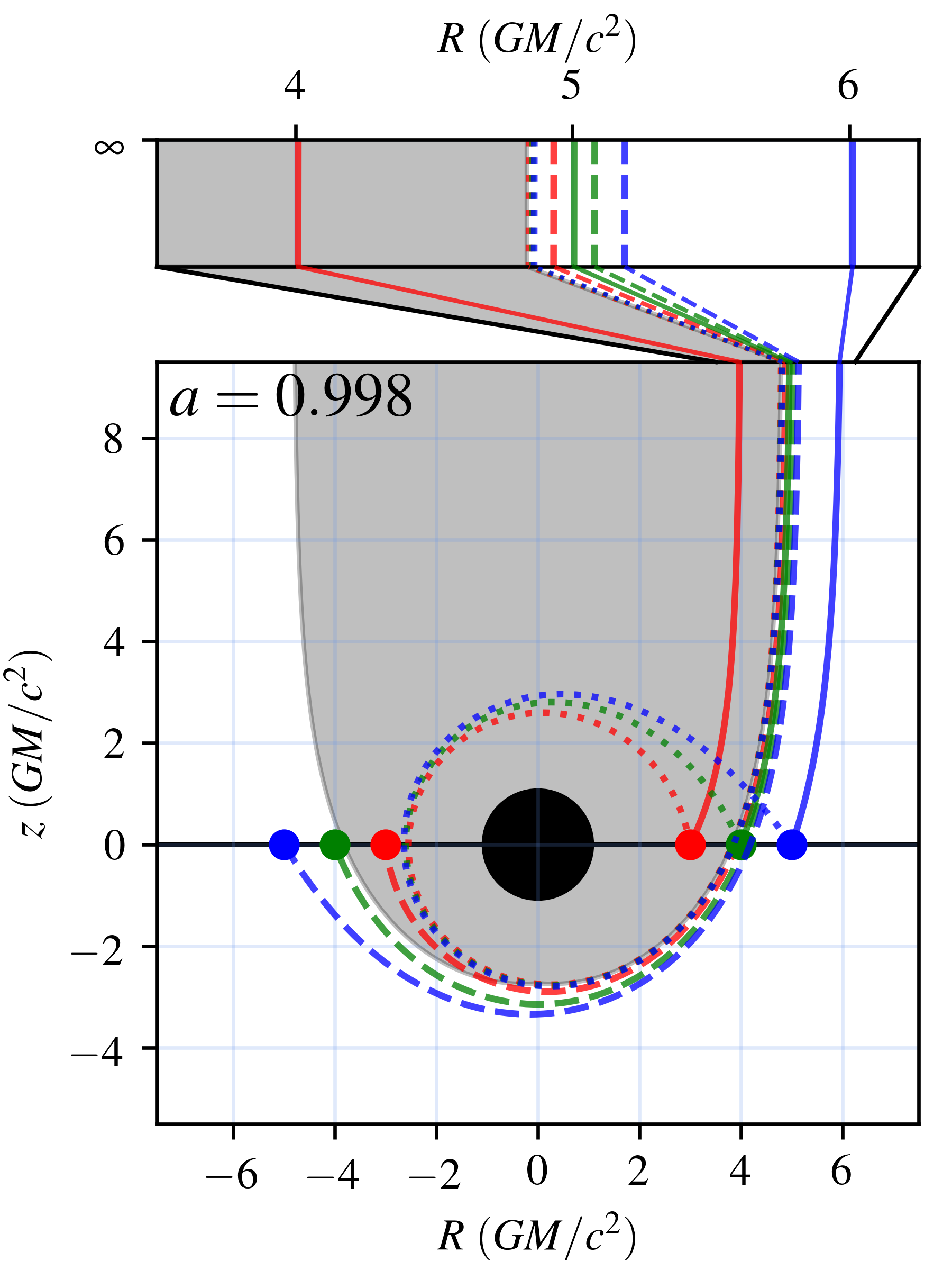}
\end{center}
\caption{Photon orbits toward a polar observer, projected into the $R\equiv r\sin\theta{\rm sgn}\phi$ and $r\cos\theta$ plane, where $r$, $\theta$, and $\phi$ are the normal Boyer-Lindquist coordinates and ${\rm sgn}\phi=\pm1$ for $|\phi|<\pi/2$ and $|\phi|>\pi/2$, respectively.  Three different emission regions in the equatorial plane (thin black line) are shown, at $r_{\rm em}=3GM/c^2$ (in red), $4GM/c^2$ (in green), and $5GM/c^2$ (in blue).  The orbits that intersect these regions for the direct emission ($n=0$), after a half orbit and thus contributing to the $n=1$ photon ring, and after a full orbit and thus contributing to the $n=2$ photon ring, are shown using solid, dashed, and dotted lines, respectively.  The bar at the top shows a zoom in on the relevant side at $z=\infty$, and thus in the image plane.  For reference, the grey region corresponds to the region within the ``shadow'', and the horizon is shown in black.}\label{fig:geometric_bias_examples}
\end{figure*}

The measurement of the \VirA mass by the Event Horizon Telescope (EHT) is predicated on the identification of the strongly lensed emission with the ring-like structures that surround the silhouette of the horizon, the black hole ``shadow'' \citep{Hilbert1917,Luminet1979,Falcke2000:shadow,BroderickLoeb2009}.  The bulk of this emission is associated with two components: (1) diffuse emission formed by those photons that follow the most direct path to the observer and (2) a ring-like structure dominated by those photons that detour around the backside of the black hole and are therefore more strongly lensed.  These components -- which we enumerate as $n=0$ and $n=1$ -- represent the first two in an infinite series of lensed images, corresponding to the primary and secondary lensed images, and now commonly referred to as ``photon rings'' associated with successively tighter photon trajectories about the black hole \citep[see, e.g.,][and references therein]{Darwin_1959, Luminet1979,Johnson_2019}.  The $n=\infty$, or asymptotic, photon ring defines the boundary of the shadow.  Examples of the $n=0$, $n=1$, and $n=2$ photon rings are shown in \autoref{fig:geometric_bias_examples}.   

When individual photon rings are spatially unresolved, the relationship between their combined emission and the size of the shadow is dependent on the location of the emission region.  Thus, the mass measurements presented in \citetalias{M87_PaperVI} rely on calibrations that employed numerical simulations of the emission region about the black hole.  The astrophysical uncertainties surrounding the structure of the emission region dominate the systematic uncertainty on the mass.

Measurements of the black hole spin are similarly complicated. Many authors have noted the modification in the shape of the shadow with spin for oblique observers, and the attendant possibility for a mass-independent spin measurement \citep{Bardeen1973,Chandrasekhar_1983,Takahashi_2004}. However, for the nearly-polar viewing geometry of \VirA, with the spin of the black hole at most $20^\circ$ to the line of sight (\citealt{Walker_2018}; \citetalias{M87_PaperV}), there is nearly no deformation in the photon ring shape, rendering such shape measurements exceedingly difficult.  Spin also produces a roughly 6\% variation in the size of the shadow for polar observers, but the $\gtrsim10\%$ precision of the current best alternative mass measurement preclude this avenue as well.  Both of these limitations prevent precision tests of general relativity at present using \VirA.

However, two fortuitous developments create an opportunity to exploit the time-variability of \VirA to overcome the dominant systematic uncertainties in measuring mass, spin, and the emission region.  The first is that novel techniques \citep{themaging,THEMIS-CODE-PAPER} and future instruments \citep{ngEHT} promise to disentangle the $n=0$ and $n=1$ photon rings, and the possibility of space-based stations, with their associated long baselines, may lead to the direct detection of the $n=2$ photon ring \citep{Johnson_2019}.

The second is that the emission within the general relativistic magnetohydrodynamic (GRMHD) models of \VirA tend to be dominated by that from the equatorial plane \citepalias{M87_PaperV}, providing a substantial simplification of the geometry of emission process.

Therefore, encoded within the photon rings and their relationships are the properties of both the emission region and the spacetime \citep[see, e.g., Fig.~10 of][]{BroderickLoeb2006}.  Here we show that with multiple measurements of the photon rings of \VirA, made when the image is dominated by emission at different radii, it is possible to break the degeneracy between where the emission occurs and how large the photon rings appear.  This procedure is fundamentally a lensing measurement, similar to those described by \citet{BroderickLoeb2006} and \citet{Tiede2020}, that leverages the evolving relationship between the different image components.  

The analysis we present here should also be understood to exist within the context of more general modeling exercises.  For \VirA, viewed very nearly along the polar axis and with the strong priors that the emission is optically thin and dominated by emission near the equatorial plane (e.g., MAD models), the procedure described here distills the key observables and their associated physical implications for the spacetime.  However, we neglect the additional information available in the large-scale flux distribution and the implications that jet kinematics has for this.  Similarly, we ignore significantly non-equatorial emission regions, e.g., from the jet funnel wall or within the jet itself.  Thus, in this sense, the analysis presented here is a subset of the joint emission region-spacetime modeling that is possible with semi-analytic analyses \citep[e.g.,][]{Tiede2020,Broderick2016,Johannsen2016,Broderick2014}, and already implemented within \themis \citep{THEMIS-CODE-PAPER}.

In \autoref{sec:origins} we describe the measurement and the underlying lensing properties that make it possible to break the degeneracies between the mass, spin, and emission region.  \autoref{sec:Mass} contains the implications for a detection of the $n=1$ or $n=2$ photon ring for mass estimates.  \autoref{sec:Maequatorial} presents the spin and mass measurements possible for a toy problem in which the emission is confined to a ring within the equatorial plane.  \autoref{sec:Maimages} explores The impact of a radial emission distribution and finite emission region height.  \autoref{sec:testingGR} describes two tests of general relativity made possible by observing multiple photon rings.  Finally, conclusions are collected in \autoref{sec:conclusions}.

\section{Origin and relationships between photon rings} \label{sec:origins}
In addition to presuming that the black hole is well described by the Kerr spacetime, we make three additional underlying assumptions:
\begin{enumerate}
\item The emission is confined to the equatorial plane.
\item The emission is axisymmetric, i.e., produced by rings, and reaches a maximum at a single radius.
\item The observer is positioned at infinity along the polar axis.
\end{enumerate}
The first two assumptions are well-justified for GRMHD models, and especially so for models in the MAD state.  The large magnetic fluxes captured by the black hole produce a high-pressure funnel region that compresses even virial accretion flows to disk heights of $h\sim 0.1 r$, where $r$ is the Boyer-Lindquist radius \citepalias[see][and references therein]{M87_PaperV}.  The accreting gas is strongly differentially rotating, rapidly shearing features into ring structures on timescales short in comparison to the accretion time.\footnote{The angular velocity in MAD models is typically more than 50\% of the appropriately relativistic Keplerian value \citep{McKinney2012}.  However, note that the details of the motion between observations is not important here.}  We will further idealize the emission as arising from a single equatorial ring; in \autoref{sec:Maimages} we show that the radius of the emission ring idealization may be replaced with the location of the emission maximum for models with extended emission.  The third assumption is well-justified by the near-polar viewing geometry of \VirA.

Given the above assumptions, the direct emission and subsequent lensed images form discrete, concentric circular rings, i.e., what are commonly called photon rings.  The observed radii of these rings are functions of the spacetime and location of the emission region, as illustrated in \autoref{fig:geometric_bias_examples}.  We begin by exploring the latter dependence.

Consider the emission from a single, azimuthally symmetric ring located in the equatorial plane with radius $r_{\rm em}$.  All emission components are constrained to connect with the same radial position in the equatorial plane, which induces a relationship between them (see \autoref{fig:geometric_bias_examples}).  The $n=0$ direct emission component is associated with those photon trajectories that connect with the emitting ring without executing any orbits about the black hole prior to reaching the observer at infinity.  The $n=1$ photon ring is composed of photons that execute a half orbit, the $n=2$ photon ring those that undergo a full orbit, etc.  The $n=\infty$ photon ring defines the boundary of the black hole shadow \citep{Bardeen1973}.

Though the radius of each photon ring depends on the black hole spin in a nontrivial fashion, the black hole mass simply acts as an overall scaling factor.  We thus have the decomposition
\begin{equation}
\begin{aligned}
\theta_{n=0} &= (GM/c^2D) \vartheta_{n=0}(a,r_{\rm em})\\
\theta_{n=1} &= (GM/c^2D) \vartheta_{n=1}(a,r_{\rm em})\\
\theta_{n=2} &= (GM/c^2D) \vartheta_{n=2}(a,r_{\rm em}),
\end{aligned} \label{eqn:MassScaling}
\end{equation}
where $D$ is the distance to the source and $GM/c^2D$ is the angular size of the gravitational radius\footnote{For reference, $GM/c^2D=3.8\pm0.4~\muas$ in \VirA \citepalias{M87_PaperVI}.}, and the $\vartheta$ are dimensionless functions that encode the dependence on $a$ and $r_{\rm em}$.  We compute these functions numerically via the procedure described in \autoref{app:vartheta}.

\begin{figure}
\begin{center}
\includegraphics[width=\columnwidth]{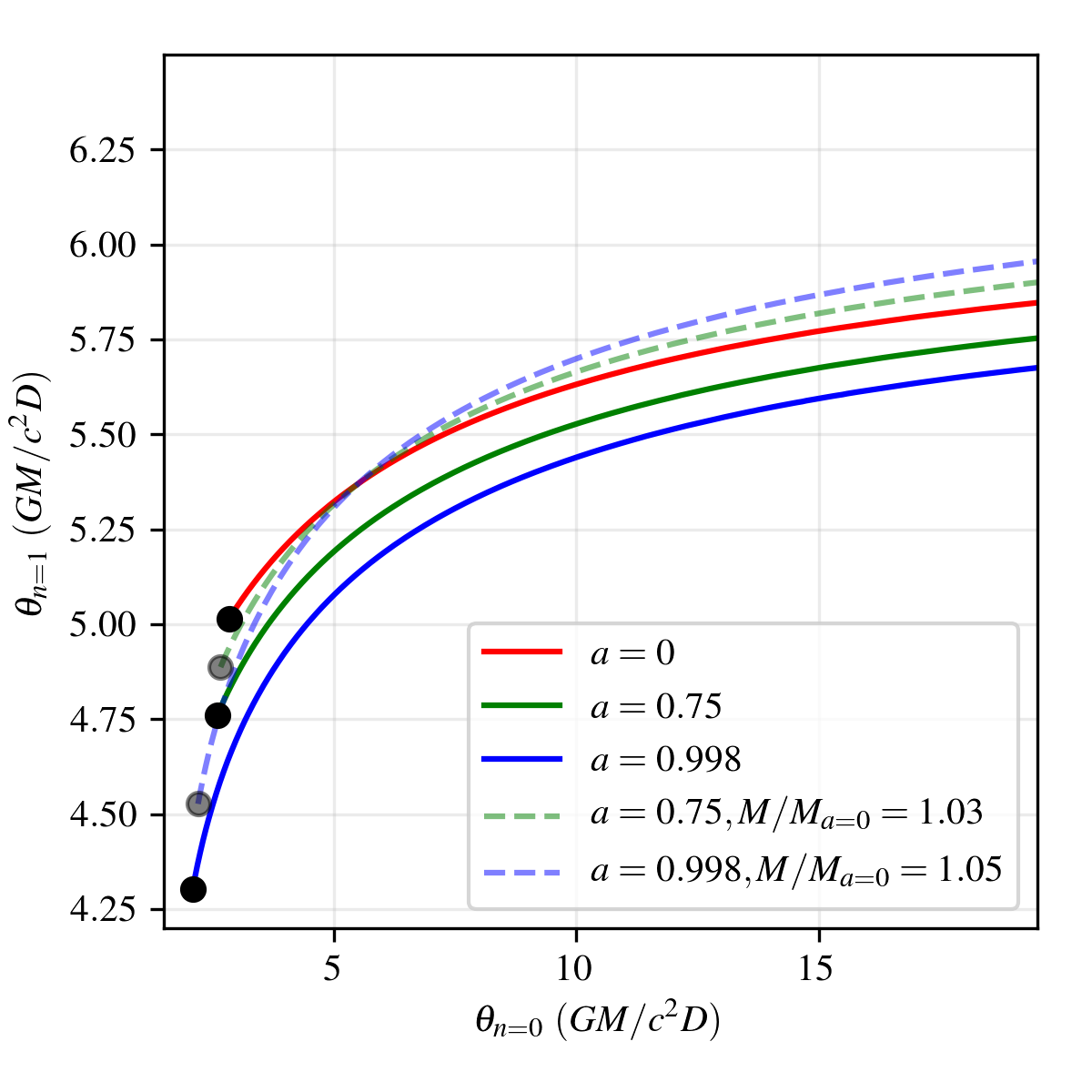}
\includegraphics[width=\columnwidth]{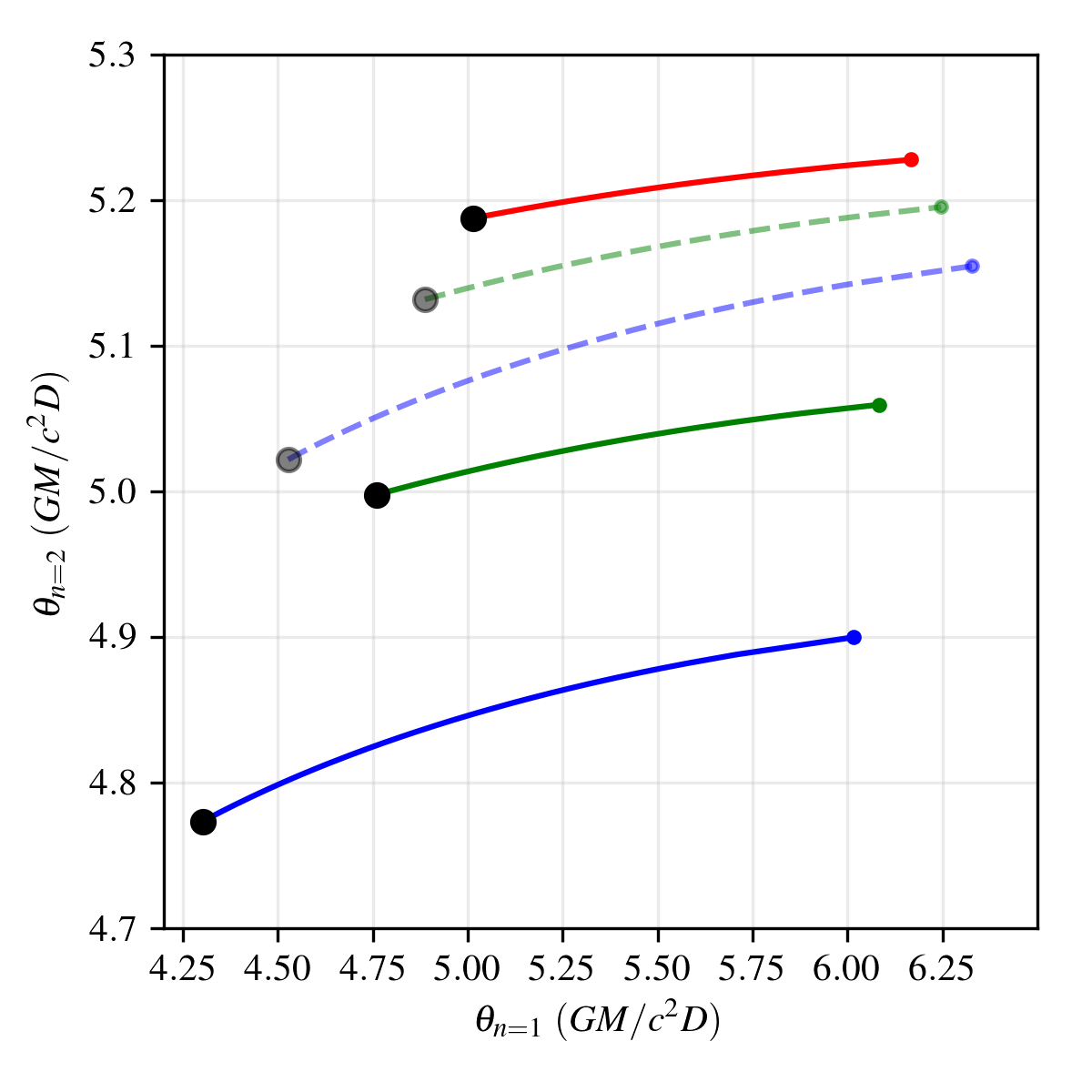}
\end{center}
\caption{Top: angular radius of the $n=1$ photon ring as a function of the angular radius of the $n=0$ photon ring for an emitting ring located at a radius $r_{\rm em}$ in the equatorial plane; the plotted curve extends from the horizon (denoted by a black circle) outward, as seen by a polar observer.  The curves associated with $a=0.75$ and $0.998$ are also shown after rescaling by a fixed mass ratio to match the $a=0$ curve at $\theta_{n=0}=5.5 GM/c^2D$.  Bottom: A similar plot for the angular sizes of the $n=2$ and $n=1$ photon rings.  The small point at the rightmost extent of the lines of the lines indicates the maximum values of $\theta_{n=1}$ and $\theta_{n=2}$.  In both panels, note the widely differing horizontal and vertical ranges.}\label{fig:thetatheta}
\end{figure}

A dependence on mass, spin, and emission location is also retained, and is perhaps more observationally accessible, through relative measurements of different-order photon rings.  The top panel of \autoref{fig:thetatheta} shows $\theta_{n=1}$ plotted against $\theta_{n=0}$ for all $r_{\text{em}}$ and $M$ and for three different fixed values of $a$.  While the horizon provides a natural minimum $r_{\rm em}$ for all orders of photon ring, $\theta_{n=0}$ is unbounded from above, and thus even after fixing $a$ (and effectively fixing $M$ by choice of units) the observed photon ring radii are a one-dimensional family traversed by $r_{\rm em}$.

These families are not, however, degenerate.  In the top panel of \autoref{fig:thetatheta} we show versions of the $a=0.75$ and $a=0.998$ curves that have had their masses rescaled in such a way as to match the $a=0$ curve at $\theta_{n=0}=5.5 GM/c^2D$ (note that the underlying $r_{\rm em}$ at the intersection point do not match).  These rescaled curves are shown by the dashed lines, and they are clearly discrepant from the $a=0$ curve for all other values of $\theta_{n=0}$.  That is, the measurement of $\theta_{n=0}$ and $\theta_{n=1}$ from emission at two different $r_{\rm em}$ is sufficient to break the degeneracy between $M$ and $a$.

The measurement of $\theta_{n=2}$ provides yet another independent constraint on $M$ and $a$.  The bottom panel of \autoref{fig:thetatheta} shows $\theta_{n=2}$ as a function of $\theta_{n=1}$ for the same three values of $a$ as in the top panel.  Because the effect of spin on the functional relationships between the different ring sizes is not a simple scaling, the $\theta_{n=1}$ and $\theta_{n=2}$ obtained by rescaling the mass do not intersect in the bottom panel of \autoref{fig:thetatheta}.  Therefore, the mass and spin may be disentangled if the $n=2$ photon ring can be measured.

In the following sections we explore the ability to leverage these over-constrained photon ring measurements to produce corresponding constraints on $M$ and $a$.

\section{$M$ from Photon Ring Detections} \label{sec:Mass}
\begin{figure*}
\begin{center}
\includegraphics[width=\textwidth]{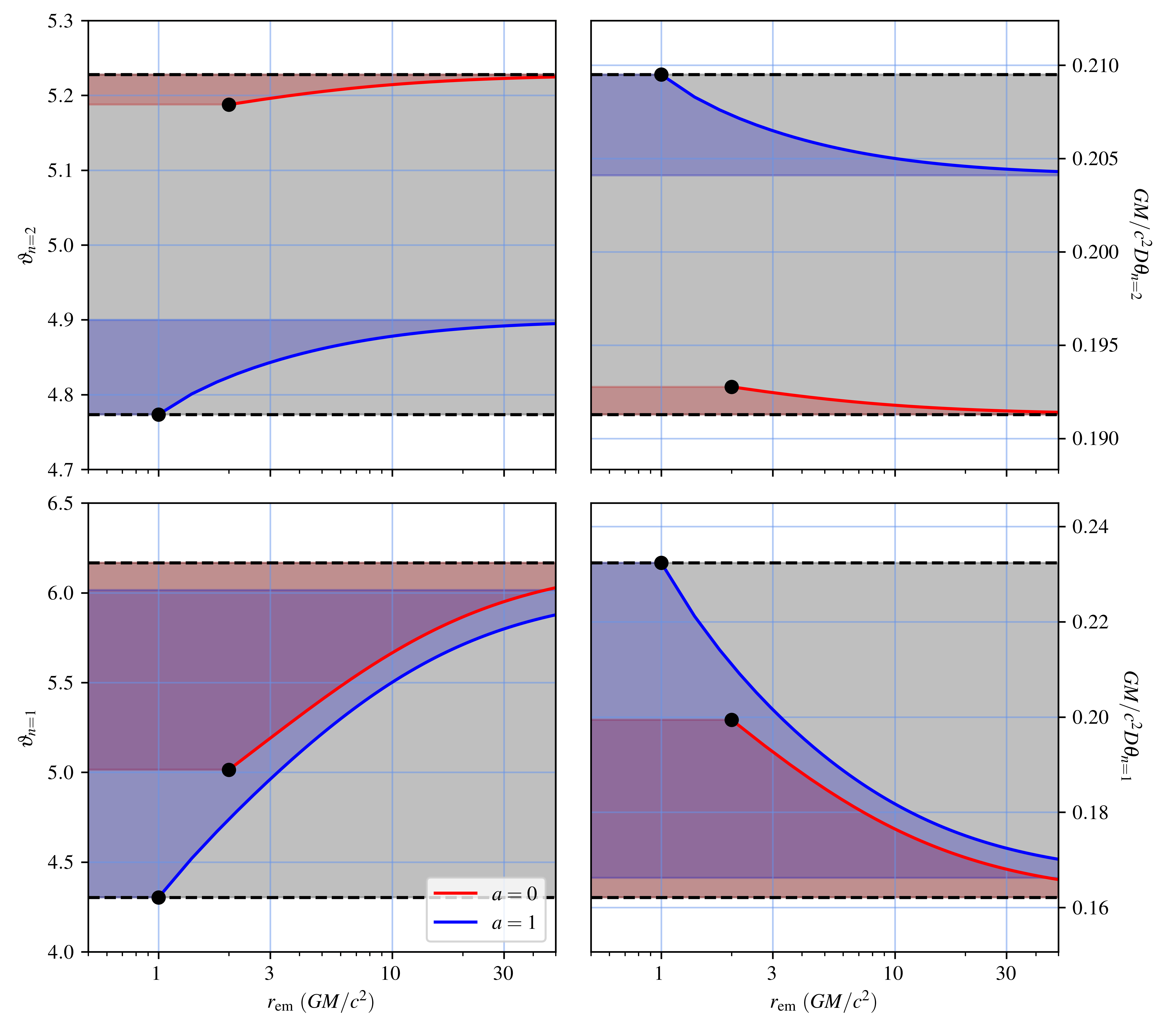}
\end{center}
\caption{Left: $\vartheta_{n=1}$ and $\vartheta_{n=2}$ as functions of the radial location of the equatorial emission, $r_{\rm em}$ extending from the horizon (black circle) outward as seen by a polar observer.  Right: the calibration coefficients necessary to convert radius measurements of the $n=1$ and $n=2$ photon rings to masses, analogous to $2/\alpha$ for the $\alpha$ coefficient in \citetalias{M87_PaperVI}.  In all panels, the range spanned by each spin is indicated by the associated colored regions, and that spanned by all spins is shown in grey and bounded by the dashed black lines.}\label{fig:Malimits}
\end{figure*}
For a given black hole mass and spin, and under the assumptions specified in \autoref{sec:origins}, the radius of a photon ring depends monotonically on the location of its emitting region within the equatorial plane, $r_{\text{em}}$.  All orders of photon ring have a minimum radius imposed by the equatorial location of the black hole horizon.  The $n=0$ photon ring radius is unbounded from above, but for photon rings beyond $n=0$ there exists a maximum radius because all contributing rays are necessarily strongly lensed.  For example, the maximum $n=1$ photon ring radius for a spin-zero black hole is simply the impact parameter for 90-degree photon scattering, corresponding to photon trajectories incoming from $r_{\text{em}} \rightarrow \infty$.  Across all values of spin and $r_{\text{em}}$, the $n=1$ photon ring is bounded on $4.30 < \vartheta_{n=1} < 6.17$ and the $n=2$ photon ring is bounded on $4.77 < \vartheta_{n=2} < 5.22$ (see \autoref{fig:Malimits}); these bounds can be compared against the minimum and maximum radii for the shadow itself of $4.83 < \vartheta_{n=\infty} < 5.20$.

Because each photon ring radius is related to the black hole mass by a simple scaling (see \autoref{eqn:MassScaling}), the boundedness of $n>0$ photon rings affords any measurement of such a photon ring a corresponding hard constraint on the black hole mass.  The size of the $n=0$ ring is unconstrained -- i.e., larger values of $r_{\text{em}}$ correspond to larger rings, without bound -- so any attempt to constrain the black hole mass using a measurement of the $n=0$ ring must rely on some prior expectation of plausible values for $r_{\text{em}}$.\footnote{This fact was highlighted by \citet{Gralla_2019} and previously explicitly addressed in \citetalias{M87_PaperVI}, where this prior was determined using a library of GRMHD simulations, from which the scaling factor relating the observed emission diameter to the size of the gravitational radius was calibrated.}  But for measurements of photon rings with $n>0$, any ``astrophysical uncertainty'' associated with an unknown or poorly-known $r_{\text{em}}$ can introduce only a finite, bounded bias on the corresponding black hole mass.  These limits are shown for the $n=1$ and $n=2$ photon rings in the right-hand panels of \autoref{fig:Malimits}, for which the mass is bounded on $0.162 \theta_{n=1} < G M/c^2 D < 0.232 \theta_{n=1}$ and $0.191 \theta_{n=2} < G M/c^2 D < 0.209 \theta_{n=2}$, respectively.\footnote{Note that the boundedness of $\theta_{n=1}$ is a generic feature of any gravitationally lensed system in which the generation of higher-order images requires a finite deflection angle.  Thus, this property is independent of the specific equatorial emission model we consider here.  The particular values of the bounds, and their implications for mass, may depend on the emission details.}  By $n=2$, the dominant systematic impacting the translation between photon ring radius and black hole mass is no longer $r_{\text{em}}$, but rather the black hole spin.

\section{$M$ and $a$ from Equatorial Emission} \label{sec:Maequatorial}
\begin{figure}
\begin{center}
\includegraphics[width=\columnwidth]{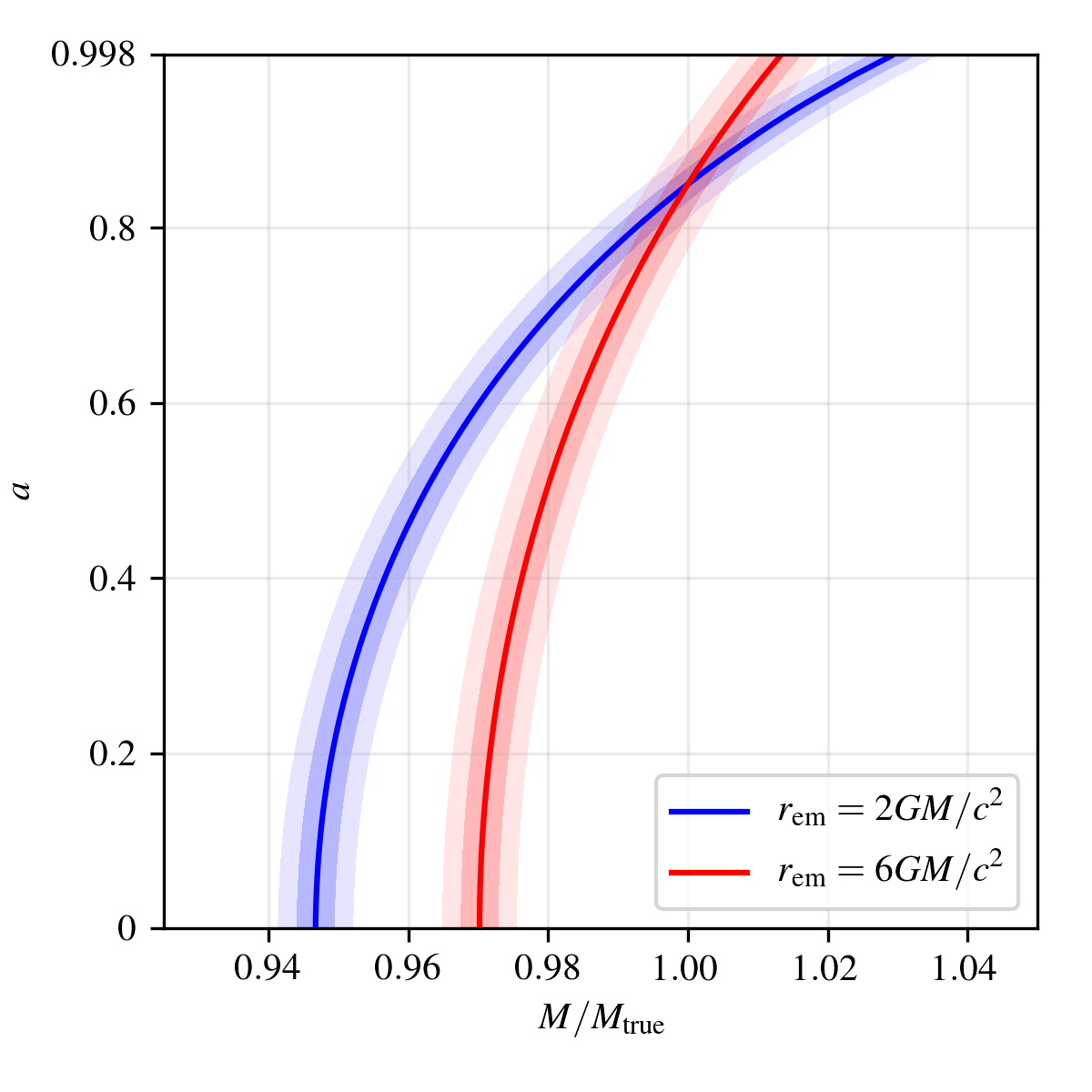}
\end{center}
\caption{Constraint on mass and spin associated with a measurement of $\theta_{n=0}$ and $\theta_{n=1}$ when the emission arises from a ring with Boyer-Lindquist radius $r_{\rm em}=2 GM/c^2$ (blue) and $r_{\rm em}=6 GM/c^2$ (red) about an $a=0.85$ black hole.  The narrow and thick shaded bands correspond to the permitted region when the errors in the angular measurements are 0.25\% and 0.5\%, respectively. Observations of emission from two or more different radii, as might be seen across multiple observing epochs, result in a unique mass and spin measurement.}\label{fig:Mademo}
\end{figure}
The horizon-scale structure of \VirA is variable.  This is seen via the presence of month-timescale ejections \citep[see, e.g.,][]{Ly2007,Walker2016,Hada2016} and the variations across the 2017 EHT campaign \citepalias{M87_PaperI,M87_PaperIV,M87_PaperVI}.  This variability is anticipated by the variable horizon-scale structures seen in GRMHD simulations \citep[see, e.g.,][]{Ripperda2020,DavisTchekhovskoy2020}.  Insofar as this variability can manifest as a change in the peak emission radius with time, it provides an opportunity to break the spin-mass degeneracy in \autoref{fig:thetatheta}.

We begin again with the idealized case in which we imagine having measurements of $\theta_{n=0}$ and $\theta_{n=1}$ from an emitting ring located at some equatorial radius $r_{\rm em}$.  A single pair $\{\theta_{n=0},\theta_{n=1}\}$ of such measurements produces a degenerate joint constraint on $M$ and $a$, traversed by (the unknown) $r_{\rm em}$; \autoref{fig:Mademo} illustrates this joint constraint for two choices of the underlying $r_{\rm em}$.  The widths of the bands in \autoref{fig:Mademo} are set almost entirely by the uncertainty in the $\theta_{n=1}$ measurement, and they are only weakly sensitive to the uncertainty of $\theta_{n=0}$.  For the purposes of \autoref{fig:Mademo} we have assumed ring radius measurement uncertainties of 0.25\% and 0.5\% for the narrow and wide bands, respectively.\footnote{For \VirA, these uncertainties correspond to $0.05~\muas$ and $0.1~\muas$, respectively.}

The degeneracy between $M$ and $a$ may be broken by measuring $\{\theta_{n=0},\theta_{n=1}\}$ pairs originating from different $r_{\rm em}$.  We will assume that these observations occur at different times, and therefore are uniquely identifiable, i.e., each $\{\theta_{n=0},\theta_{n=1}\}$ pair may be determined.  However, should multiple emission rings be present, corresponding to multiple emission maxima (see \autoref{sec:Maimages}), these may produce sufficient information to break the $M$-$a$ degeneracy in a single observation epoch.

An example of this degeneracy breaking is illustrated in \autoref{fig:Mademo}, which shows two bands corresponding to photon rings emitted from Boyer-Lindquist radii of $2GM/c^2$ and $6GM/c^2$ around a black hole with $a=0.85$\footnote{Both rings are safely outside of the prograde photon orbit radius, which for $a=0.85$ is $1.6938GM/c^2$.  The smaller ring is within the equatorial prograde innermost circular orbit (ISCO), which is located at $2.6321GM/c^2$.  However, neither GRMHD simulations nor radiatively-inefficient accretion flow (RIAF) calculations indicate significant features at the ISCO for the kinds of accretion flows believed to be relevant for \VirA.  Nevertheless, our qualitative results are independent of the particular values of $r_{\rm em}$ chosen.}.  The overlap region between these two bands covers the true mass and spin values.  The degree to which two such measurements produce useful mass or spin constraints depends primarily on the uncertainties in the measurements of $\theta_{n=1}$.  It depends further on the difference in $r_{\rm em}$ between the two observations; two measurements with the same emission location are naturally degenerate with each other and provide no leverage with which to separate $M$ and $a$.

\begin{figure}
\begin{center}
\includegraphics[width=\columnwidth]{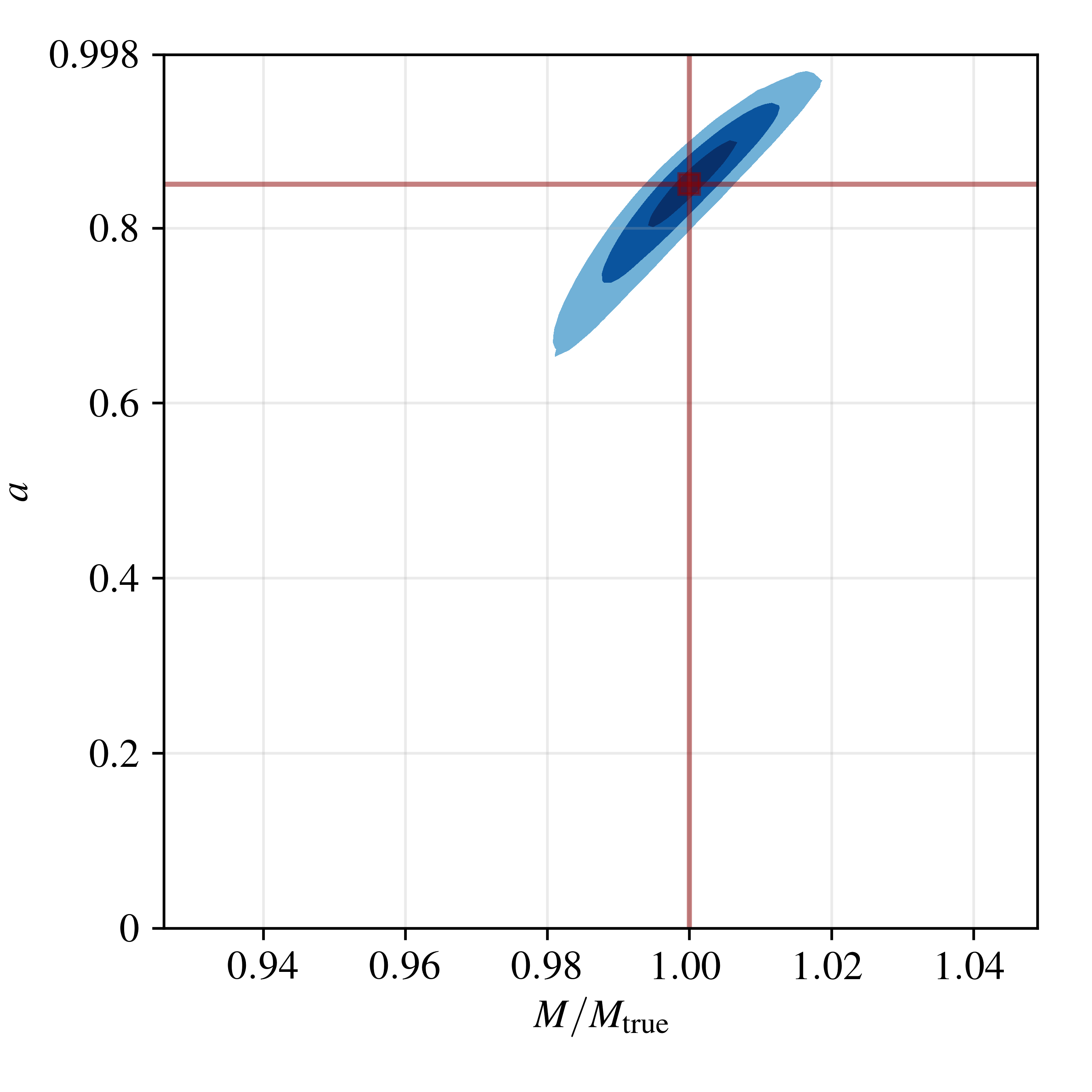}
\end{center}
\caption{Joint posterior on mass and spin associated with a pair of simulated measurements of $\theta_{n=0}$ and $\theta_{n=1}$ when the emission arises from a ring with Boyer-Lindquist radius $r_{\rm em}=2 GM/c^2$ and, later, $r_{\rm em}=6 GM/c^2$ about an $a=0.85$ black hole.  Contours indicate cumulative 50\%, 90\%, and 99\% regions.  Angular measurements are assumed to have a precision of 0.25\%.  The truth values are indicated in dark red.}\label{fig:Mapostdemo}
\end{figure}

A practical demonstration of the constraints on mass and spin that can be obtained from a pair of $\{\theta_{n=0},\theta_{n=1}\}$ measurements with 0.25\% Gaussian uncertainties is shown in \autoref{fig:Mapostdemo}.  From these synthetic measurements we constructed a Gaussian likelihood and sampled the corresponding posterior using the ensemble Markov Chain Monte Carlo (MCMC) method provided by the \texttt{emcee} python package \citep{emcee2013}.  We use 64 independent walkers and run for $10^5$ steps, discarding the first half of each chain.  Explorations with fewer walkers and steps indicate that by this time the MCMC chains are well converged.  We see that the posterior matches well with the general shape of the overlap region in \autoref{fig:Mademo}.

\begin{figure}
\begin{center}
\includegraphics[width=\columnwidth]{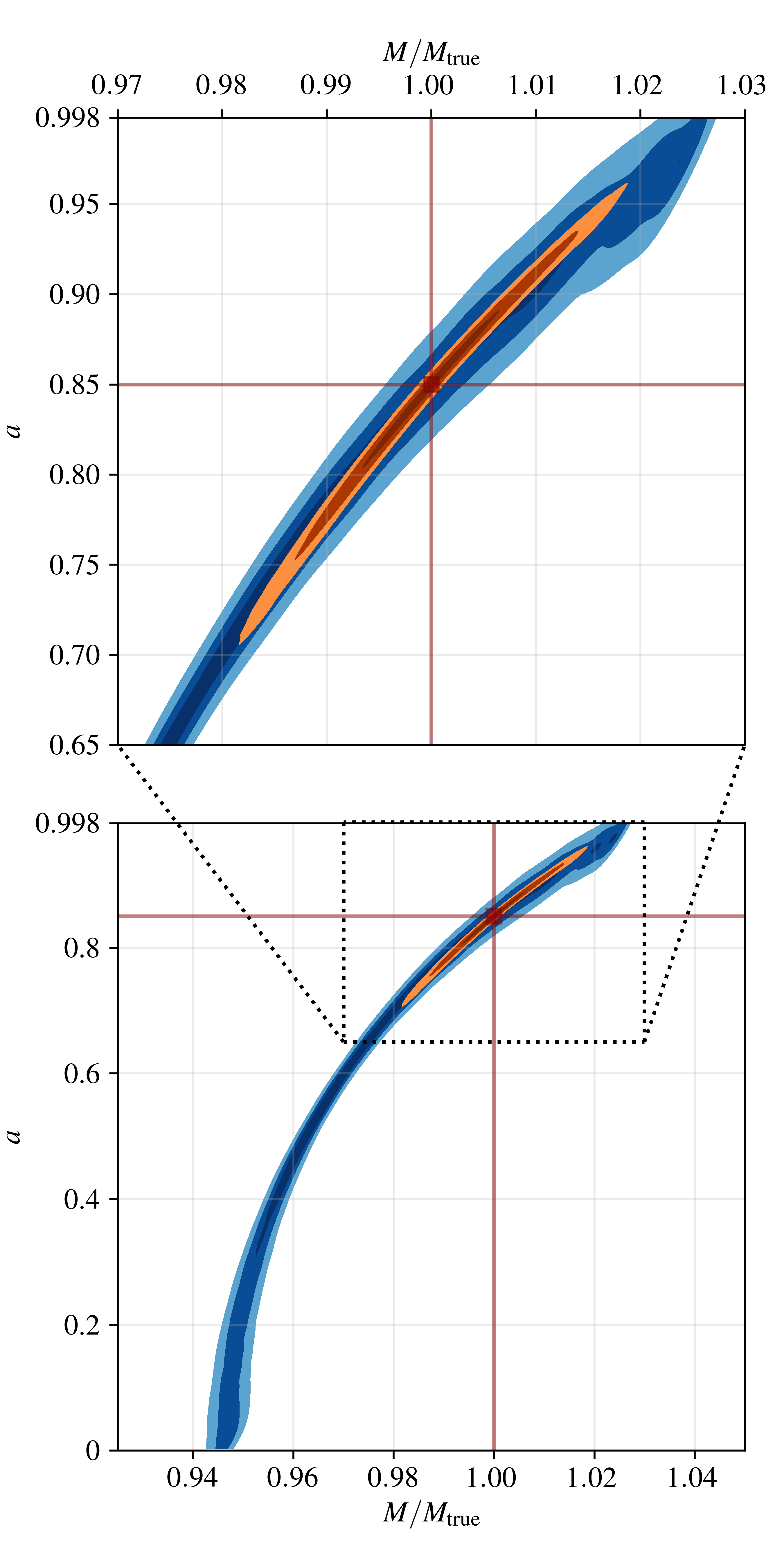}
\end{center}
\caption{Joint posterior on mass and spin associated with a single simulated measurement of $\theta_{n=0}$, $\theta_{n=1}$, and $\theta_{n=2}$ when the emission arises from a ring with Boyer-Lindquist radius $r_{\rm em}=2 GM/c^2$ about an $a=0.85$ black hole.  Shown in blue is the posterior from only the $\theta_{n=0}$ and $\theta_{n=1}$ measurements; shown in orange is the posterior when $\theta_{n=2}$ is included (magnified above).   Contours indicate cumulative 50\%, 90\%, and 99\% regions.  Angular measurements are assumed to have a precision of 0.05\%.  Truth values are indicated in dark red.}\label{fig:Ma012postdemo}
\end{figure}

If $\theta_2$ can be measured in addition to $\theta_1$ and $\theta_0$, then emission from a single $r_{\text{em}}$ can be used to break the degeneracy between $a$ and $M$.  An example such joint constraint is shown in \autoref{fig:Ma012postdemo}.  The degree to which the degeneracy is broken is very sensitive to the precision with which $\theta_2$ can be measured.  For the purposes of \autoref{fig:Ma012postdemo} we have assumed a measurement precision of 0.05\%.\footnote{For \VirA, this precision corresponds to $0.005~\muas$.}

We note that because the mass manifests as an overall, omnipresent scale, the measurements of all other quantities are dependent only on {\em relative} angular measurements.  Therefore, the black hole spin and location of the emission regions (measured in gravitational radii) can be estimated with much greater accuracies than would be implied by the systematic uncertainties associated with $M$, e.g., the uncertain distance to \VirA.

\section{$M$ and $a$ from Imaging Simulations} \label{sec:Maimages}
The demonstrations in the previous section are highly idealized, associated with a single ring-like emission region confined to the equatorial plane.
More commonly the emission is expected to be distributed radially throughout the equatorial plane and vertically off of the equatorial plane, and the emission within the photon rings will be modestly optically thick.  Even subject to the remaining assumption that the viewer is polar, these present the possibility of biases in the ring positions and corresponding spacetime-parameter estimates.

\begin{figure}
\begin{center}
\includegraphics[width=\columnwidth]{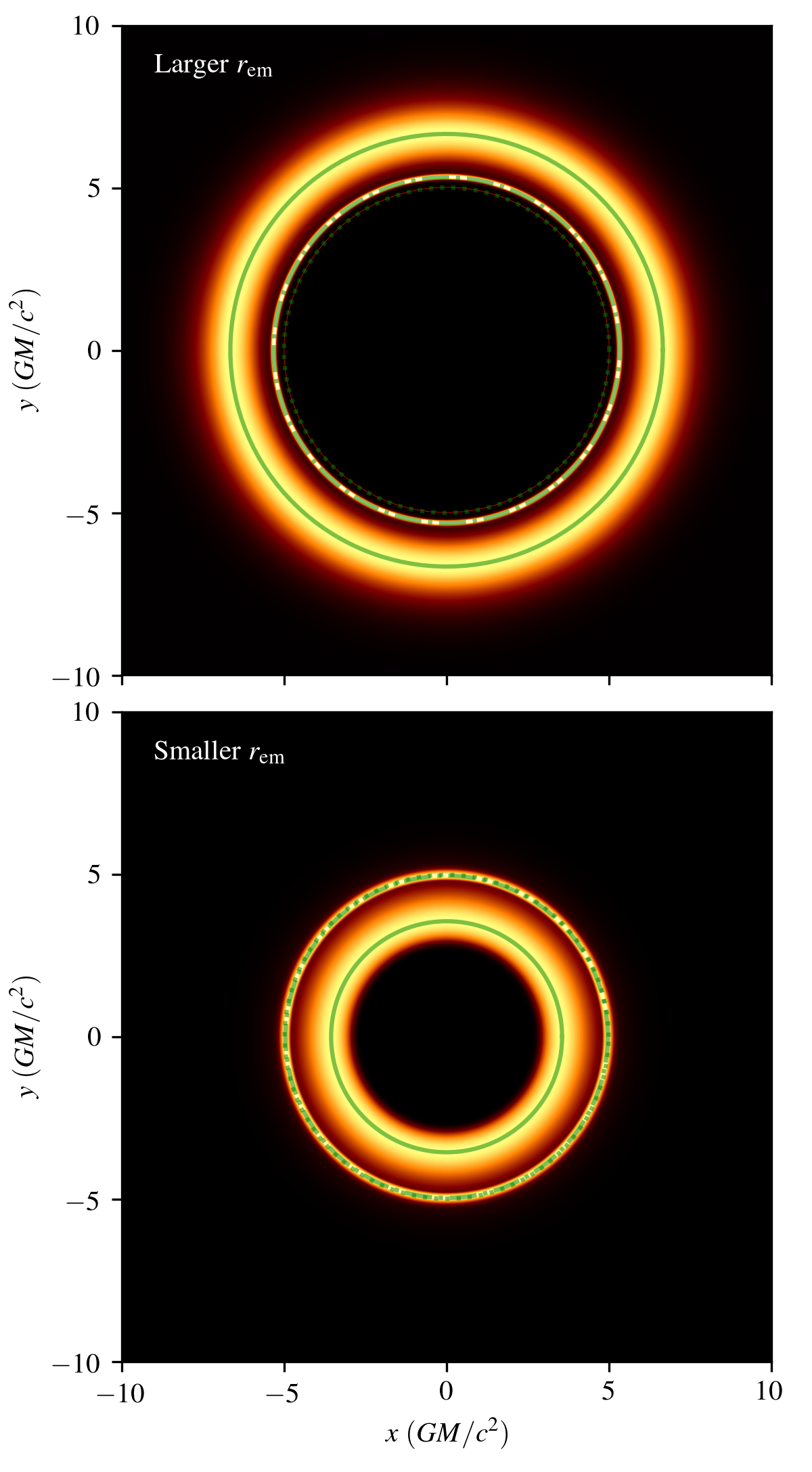}
\end{center}
\caption{Images of slim accretion flow models ($h/r=0.1$) around an $a=0.8$ black hole as seen by a polar observer with different inner-disk edges.  The solid, dash-dot, and dotted green circles indicate the locations of the peaks of the emission within the $n=0$, $n=1$, and $n=2$ photon rings.  The parameters of the emission models are listed in \autoref{tab:riaf_models}.}
\label{fig:riaf_images}
\end{figure}

To assess these we consider two images from slim radiatively inefficient accretion flow simulations about an $a=0.8$ black hole, similar to those described in \citet{BroderickLoeb2006b}.  The mass, distance and total flux are chosen to be consistent with those of \VirA \citepalias[see, e.g.,][]{M87_PaperVI}.  The emission is generated via synchrotron from thermal and nonthermal electron populations.  The temperature is virial and both electron populations are power laws in radius and Gaussian in height, with $h=0.1r$, roughly consistent with the MAD models in GRMHD simulations (\citealt{Narayan2003},\citealt{McKinney2012}, \citetalias{M87_PaperV}).  In addition, an internal cutoff in the electron densities was imposed to modify the typical location of the emission region.  That is, for each electron species $s$, the density is assumed to be
\begin{equation}
    n_s \propto r^{-\alpha_s} e^{-z^2/2h^2}
    \begin{cases}
        e^{-(r-r_{\rm cut})^2/2}
        & r<r_{\rm cut}\\
        1
        & \text{otherwise.}
    \end{cases}
\end{equation}
The orbital motion is assumed to be on circular geodesics outside of the innermost stable circular orbit, and on plunging ballistic inside with the specific energy and specific angular momentum set by that at the ISCO.  The magnetic field was set by choosing a fixed plasma $\beta=10$.  The resulting images are shown in \autoref{fig:riaf_images} when $r_{\rm cut}=6GM/c^2$ (top) and $2GM/c^2$ (bottom).  The radiative transfer computation includes synchrotron self-absorption and follows the full complement of Stokes parameters \citep{Broderick2004}.  Optical depths at the peak of the $n=0$ and $n=1$ photon rings are $\tau\approx1$, and in the $n=2$ photon ring they can reach as high as $\tau\approx2$.

\begin{figure}
\begin{center}
\includegraphics[width=\columnwidth]{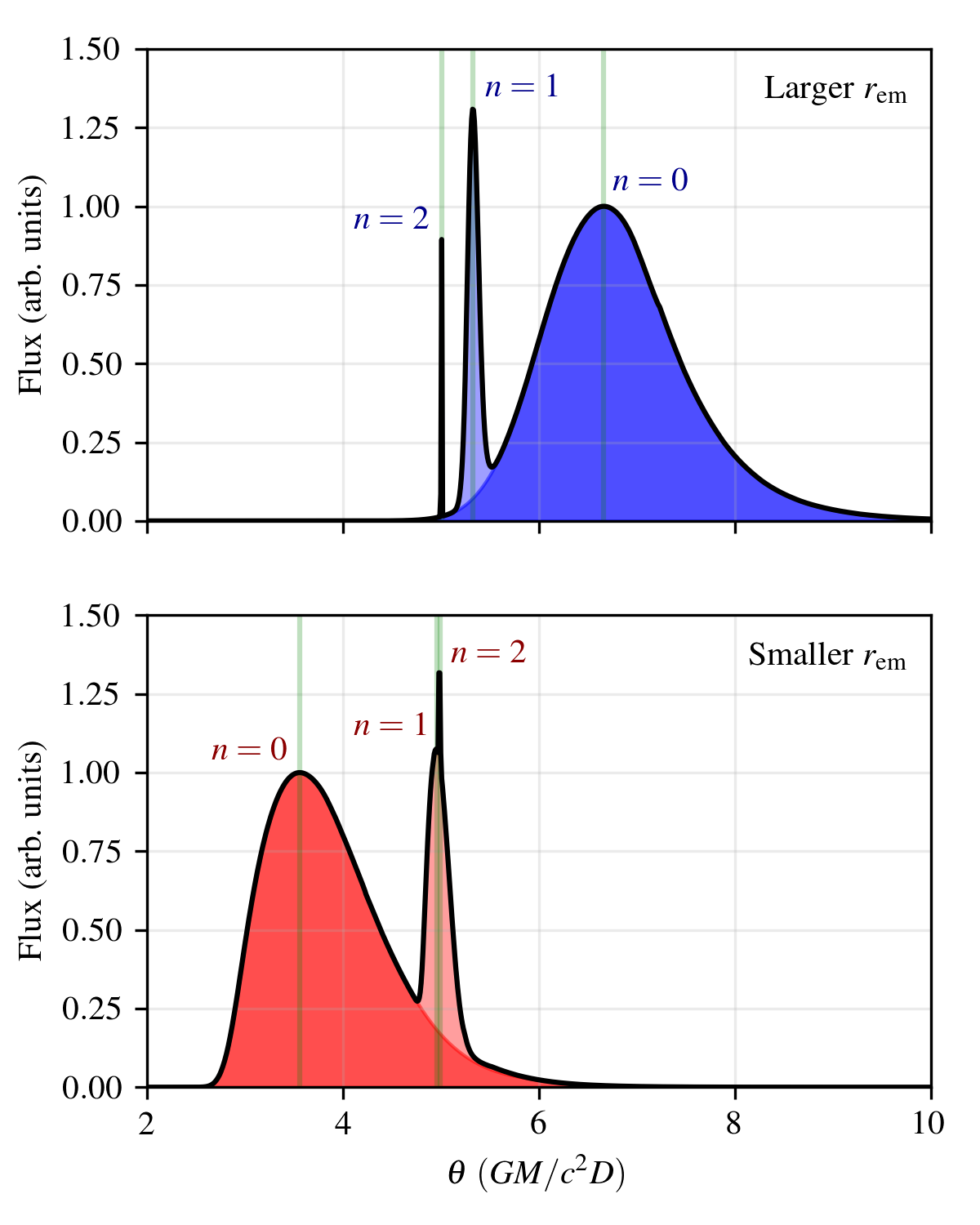}
\end{center}
\caption{Radial profiles from the slim accretion flow models shown in \autoref{fig:riaf_images}.  The top and bottom panels show profiles associated with accretion flows with large and small emission regions, respectively. The peaks of the $n=0$, $n=1$, and $n=2$ photon rings are indicated by the green lines.}\label{fig:riaf_profiles}
\end{figure}

The radial flux profiles, shown in \autoref{fig:riaf_profiles}, clearly illustrate the distributed emission.  It is asymmetric about the $n=0$ peak, a direct consequence of the asymmetric radial distribution of the emitting electrons.  In addition, the subsequent higher-order photon rings are superimposed upon the emission of the $n=0$, and in the case of the smaller radial cutoff $n=1$, photon ring(s).

\begin{deluxetable}{cccccc}
\tablecaption{Ring Radii Estimates \label{tab:riaf_models}}
\tablehead{
\colhead{$r_{\rm cut}$}\tablenotemark{a} & 
\colhead{$r_{\rm em}$}\tablenotemark{b} & 
\colhead{$n$} & 
\colhead{Meas.\ $\theta_{n}$}\tablenotemark{c} &
\colhead{Equat.\ $\theta_{n}$}\tablenotemark{d} &
\colhead{Frac.\ error}
}
\startdata
 6.0 & 5.6833 & 0 & 6.6602 & 6.6602 &  0 \\
     &        & 1 & 5.3253 & 5.3265 &  0.0002\\
     &        & 2 & 5.0047 & 5.0049 & $<5\times10^{-5}$\\
 \hline
 2.0 & 2.5940 & 0 & 3.5566 & 3.5566 &  0 \\
     &        & 1 & 4.9555 & 4.9590 &  0.0007\\
     &        & 2 & 4.9837 & 4.9838 &  $<3\times10^{-5}$\\
\enddata
\tablenotetext{a}{Measured in units of $GM/c^2$.}
\tablenotetext{b}{Emission radius estimate from $\theta_{n=0}$ in units $GM/c^2$.  For comparison, the photon orbit radius and ISCO for $a=0.8$ is $1.8111GM/c^2$ and $2.9066GM/c^2$, respectively.}
\tablenotetext{c}{Angular radius of the peak of the $n$th-order photon ring in the intensity profile in units of $GM/c^2D$.}
\tablenotetext{d}{Angular radius of $n$th-order photon ring from a ring of emission located at $r_{\rm em}$ in units of $GM/c^2D$.}
\end{deluxetable}

We identify the location of the photon ring peaks from the radial flux profiles via an iterative procedure that begins with $n=0$ and successively attempts to remove the lowest order photon ring emission.  This effectively removes the biases in the peak locations due to the overlap of the various image components.  More detail on this procedure is presented in \autoref{app:photon_ring_id}.  The resulting estimates of the $n=0$, $n=1$, and $n=2$ photon ring positions are listed in \autoref{tab:riaf_models}.

For comparison, we also determine a set of idealized photon-ring position estimates.  We do this by determining the equatorial emission radius associated with the measured $\theta_{n=0}$ (listed in the second column of \autoref{tab:riaf_models}) and then infer the locations of the remaining photon rings assuming all of the emission is confined to a ring in the equatorial plane with this radius.  These estimates and the fractional difference are also listed in the fifth and sixth column of \autoref{tab:riaf_models}.  The identification of the flux profile peaks with the position of the photon rings from an equatorial emission ring at some radius appears to be an excellent approximation.

\begin{figure}
\begin{center}
\includegraphics[width=\columnwidth]{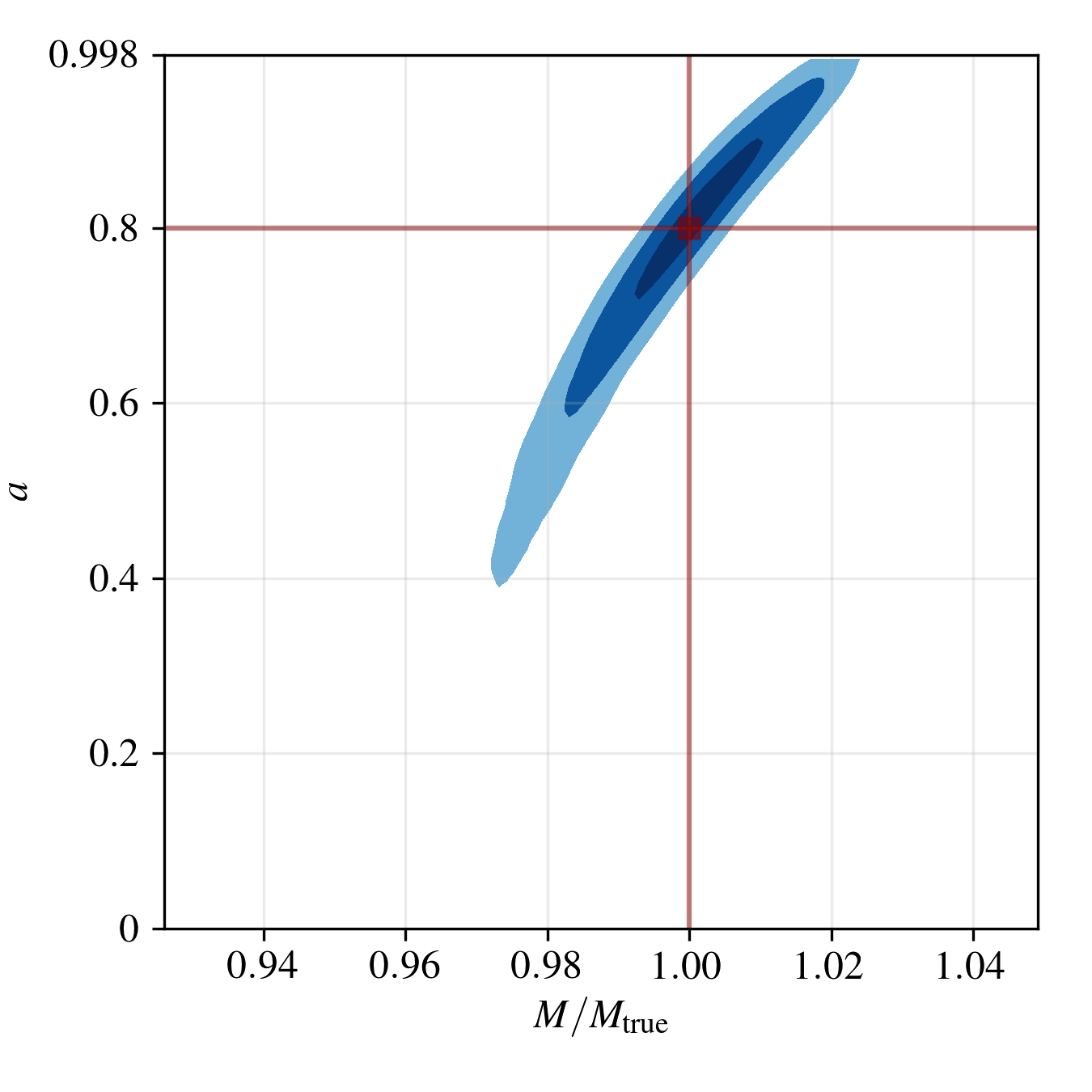}
\end{center}
\caption{Joint posterior on mass and spin associated with a pair of measurements of $\theta_{n=0}$ and $\theta_{n=1}$ from the simulated accretion flow images shown in \autoref{fig:riaf_images}.  Contours indicate cumulative 50\%, 90\%, and 99\% regions.  Angular measurements were assumed to have an accuracy of 0.25\%.  The truth values are indicated in dark red. }\label{fig:riaf_2epost}
\end{figure}

\begin{figure}
\begin{center}
\includegraphics[width=\columnwidth]{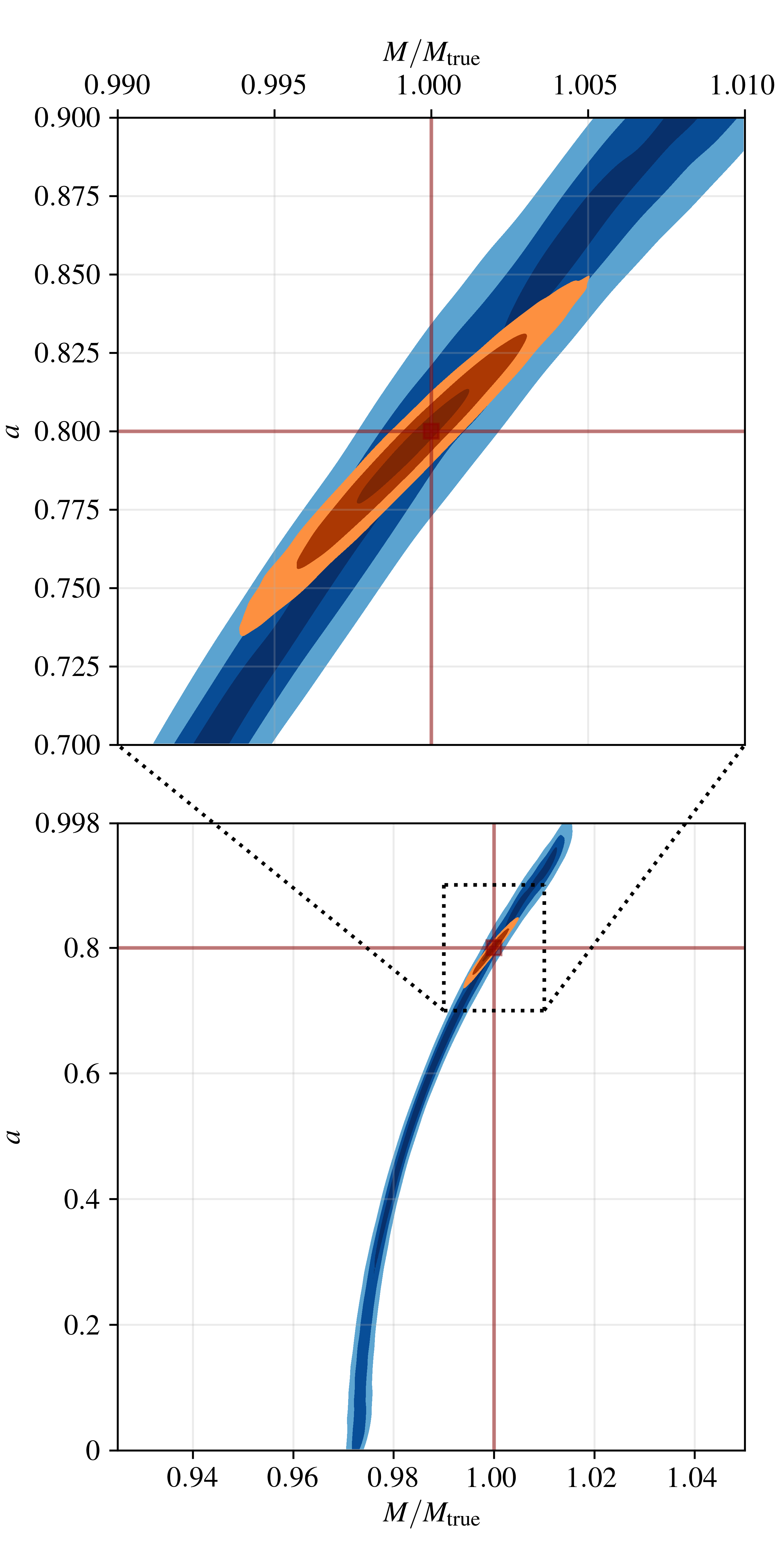}
\end{center}
\caption{Joint posterior on mass and spin associated with a single simulated measurement of $\theta_{n=0}$, $\theta_{n=1}$, and $\theta_{n=2}$ from the simulated accretion flow images shown in \autoref{fig:riaf_images} and \autoref{fig:riaf_profiles}.  Shown in blue is the posterior from only the $\theta_{n=0}$ and $\theta_{n=1}$ measurements; shown in orange is the posterior when $\theta_{n=2}$ is included (magnified above).   Contours indicate cumulative 50\%, 90\%, and 99\% regions.  Angular measurements were assumed to be measured with an accuracy of 0.05\%.  Truth values are indicated in dark red.}\label{fig:riaf_1epost}
\end{figure}

We perform the same demonstration analyses as in \autoref{sec:Maequatorial}.  The implication for mass and spin of measuring $(\theta_{n=0},\theta_{n=1})$ to 0.25\% on two epochs with different $r_{\rm em}$ is shown in \autoref{fig:riaf_2epost}.  This is comparable to \autoref{fig:Mapostdemo}, albeit at slightly different choices of the emission radii.
When $(\theta_{n=0},\theta_{n=1},\theta_{n=2})$ are measured to 0.05\% in a single epoch, the constraints on mass and spin are shown in \autoref{fig:riaf_1epost}.

For the simplified simulated images explored in this section, we have ignored a number of additional potential complications.  Our simulations reach only modest optical depths ($\tau\lesssim2$) in the images; much higher optical depths may modify the relationship between $\theta_{n=1,2}$ and $\theta_{n=0}$.  We have also considered only face-on systems; nonzero inclination and the consequent breaking of axisymmetry would make the identification of the ring radii more difficult.  Both of these limitations are most naturally addressed through more comprehensive forward-modeling of the image structure \citep[e.g.,][]{Broderick2016}.  Nevertheless, it is clear that emission regions extended in radius and modestly in height do not provide a significant impediment to spin and mass measurements.

\section{Testing General Relativity} \label{sec:testingGR}
\begin{figure*}
\begin{center}
\includegraphics[width=\textwidth]{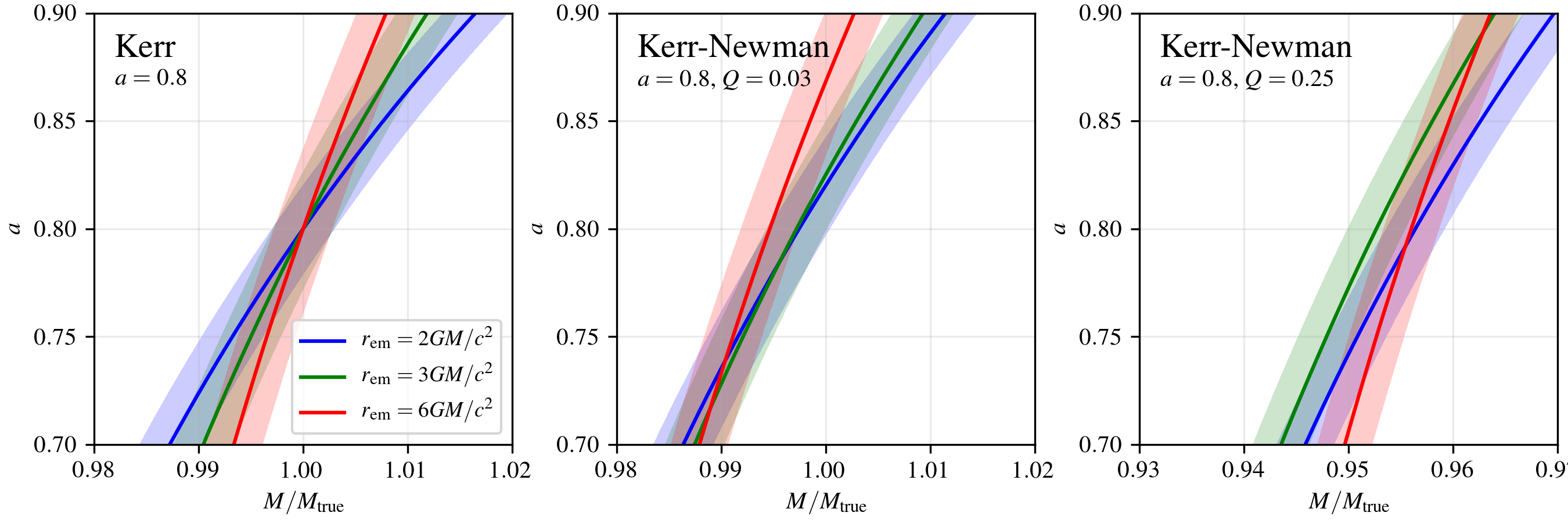}
\end{center}
\caption{Comparison of mass-spin contours from observations of $\theta_{n=0}$ and $\theta_{n=1}$ made assuming Kerr (left) and Kerr-Newman (right) spacetimes over three epochs with differing $r_{\rm em}$.  Bands indicate a 0.25\% uncertainty in the measurement of $\theta_{n=1}$.  Truth values for the spin (both) and charge (Kerr-Newman) are listed in the panels.}\label{fig:gr3e-01}
\end{figure*}
\begin{figure*}
\begin{center}
\includegraphics[width=\textwidth]{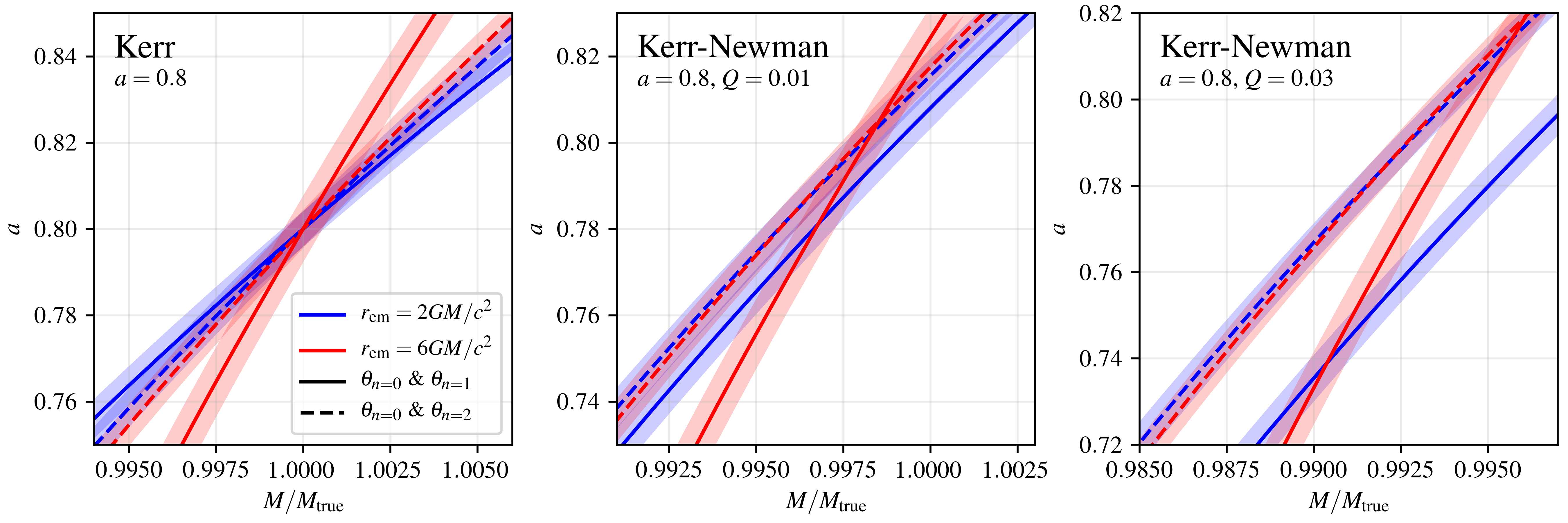}
\end{center}
\caption{Comparison of mass-spin contours from observations of $\theta_{n=0}$, $\theta_{n=1}$, and $\theta_{n=2}$ made assuming Kerr (left) and Kerr-Newman (right) spacetimes over two epochs with differing $r_{\rm em}$.  Solid and dashed contours show the $\theta_{n=0}$-$\theta_{n=1}$ and $\theta_{n=0}$-$\theta_{n=2}$ constraints, respectively.  The bands indicate a 0.05\% uncertainty on the measurements of $\theta_{n=1}$ and $\theta_{n=2}$.  Truth values for the spin (both) and charge (Kerr-Newman) are listed in the panels.}\label{fig:gr2e-012}
\end{figure*}
Careful measurement of the asymptotic photon ring provides a number of opportunities to probe general relativity more generally \citep{Takahashi_2004,JP2010,Psaltis2015,Johannsen2016,Medeiros2020}.  These have been applied to the existing \VirA EHT images (\citealt{Psaltis2020}, though see \citealt{Kocherlakota2021,Volkel2021}).  These methods require either a careful measurement of the photon ring shape, additional ancillary measurements of mass and spin, or are otherwise limited in their implications.  Here we describe a possible precision constraint produced from the radio interferometry alone.

Where two independent measurements of $\{\theta_{n=0},\theta_{n=1}\}$ from distinct emission locations during different epochs provide a measurement of mass and spin, any additional constraint in the $M$-$a$ plane would render the problem over-constrained and would therefore permit a test of general relativity.  While this supplementary measurement may be the result of ancillary observations (e.g., mass estimates from stellar orbits), it need not be: a measurement during a third epoch with a differing $r_{\rm em}$ is sufficient.

In such a test, the constancy of the underlying spacetime parameters, spin and mass, implies that all constraints in the $M$-$a$ plane must be  consistent, i.e., all bands in the $M$-$a$ plane must intersect at a single point as long as the spacetime is described by the Kerr metric.  This test is conceptually identical to the tests of general relativity presented in pulsar studies \citep[see, e.g.,][]{GRPulsar}, where the various constraints on the masses of the members of the binary are over-constrained by various metric-dependent observables.  An example multi-measurement test is shown in \autoref{fig:gr3e-01} (left panel), which shows three hypothetical observations at emission radii ranging from $2 GM/c^2$ to $6 GM/c^2$ for a black hole with $a=0.80$.  As anticipated, all bands converge at a single point, both improving the estimates of $M$ and $a$ and demonstrating the self-consistency of the Kerr metric.

In contrast, we show two additional simulated comparisons, with  $\{\theta_{n=0},\theta_{n=1}\}$ measured from images in a manner identical to that in \autoref{sec:Maimages} but generated in a Kerr-Newman spacetime with charges $Q=0.03$ and $Q=0.25$ in the center and right panels of \autoref{fig:gr3e-01}, respectively.  In practice, Kerr-Newman is a proxy for any non-Kerr metric that permits a well-defined and simple demonstration of the impact of deviations from Kerr of the photon ring properties; practical applications to specific metrics would require constructing the low-order analogs of the Kerr photon rings.  Nevertheless, we note that $Q=0.25$ could be excluded by observations of similar quality to those that can be performed in the near future.

If the $n=2$ photon ring can be detected and $\theta_{n=2}$ measured, it is possible to test general relativity in a pair of observations.  Moreover, the detection of the $n=2$ photon ring implies a significantly improved precision in the measurement of $\theta_{n=2}$ over that required to measure $\theta_{n=1}$.  Each pair of photon ring radii implies a joint constraint in the $M$-$a$ plane, yielding three such bands for each measurement.  However, only two of these constraints are typically distinct; those arising from $\theta_{n=1}$ and $\theta_{n=2}$ are nearly degenerate with those due to $\theta_{n=0}$ and $\theta_{n=2}$.  A second epoch of observations during a period with a differing $r_{\rm em}$ produces a second, independent measurement of $a$-$M$, associated with two additional constraints on $a$-$M$.  As before, the constancy of the spacetime requires that these are consistent with the prior values, i.e., all bands must cross at a single location.  

\autoref{fig:gr2e-012} shows example constraints in the $M$-$a$ plane when $\{\theta_{n=0},\theta_{n=1},\theta_{n=2}\}$ are measured from images.  As before, when the spacetime is Kerr, all constraints intersect at a single mass and spin ($a=0.8$).  Even small deviations ($Q=0.01$) could be excluded in this instance, though the precision with which this could be done depends sensitively on the precision of the measurements of $\theta_{n=2}$ and $\theta_{n=1}$.    

To lowest order, the additional spacetime structure encoded in $Q$ modifies the relationship between spin and the quadrupole moment that the methods presented here exploit. These lensing-based signatures are, in essence, what semi-analytic RIAF models are probing \citep[see, e.g.,][]{Broderick2014,Johannsen2016}. Methods relying on the shadow {\it shape} invariably face the unfortunate effect of partial cancellation of spin and quadrupole effects in the lensed image seen at infinity, leaving only mild sensitivity on black hole spin and, by extension, deviations from Kerr.  Like shape measurements, tests based on multiple photon ring size measurements are independent of the over-all mass scale, simplifying their comparison.  In addition, the impact of differing systematic uncertainties is limited, because they employ only high-resolution radio imaging data, instead of relying on ancillary measurements (such as stellar dynamical masses) --- the precision of the gravitational test depends solely on the precision of the ring size measurements.

While we focus on \VirA here, we note that in systems with significant auxiliary independent measurements of mass and spin, it is possible to supplement the constraints presented here directly.  This is the case, e.g., for \SgrA for which independent high-precision mass measurements exist \citep{Boehle2016,GRAVITY2019}.  Such posterior distributions are narrow in mass but (currently) uninformative in spin and are therefore partially complementary to the contours discussed here. As a result the mass measurements can be improved and help to further constrain spin and test general relativity beyond what is shown in \autoref{fig:gr3e-01} and \autoref{fig:gr2e-012}.  The situation would further improve with the detection of a pulsar orbiting a supermassive black hole where mass and perhaps spin could be reconstructed with better accuracy than \citet{Boehle2016,GRAVITY2019} \citep{Wex2020}.

\section{Conclusions} \label{sec:conclusions}
In this paper we have presented a number of idealized analyses demonstrating that relative size measurements for different orders of photon ring permits can be used to place constraints on black hole mass and spin {\em without} having to resolve the shape of the asymptotic photon ring.  Specifically, we have focused on the opportunities afforded by multi-epoch measurements of $\theta_{n=0}$ and $\theta_{n=1}$ --- such as are expected to be possible using multiple EHT observing campaigns --- or from a single-epoch measurement of $\theta_{n=0}$, $\theta_{n=1}$, and $\theta_{n=2}$, such as may be obtainable using future space missions.  Both of these methods leverage the relationship between the sizes of the different-order photon rings to eliminate the degeneracy that is otherwise present between the location of the emitting region (e.g., orbital radius of the emitting material) and the spacetime properties (i.e., black hole mass and spin).

These are fundamentally lensing measurements, and are similar in spirit to simultaneous reconstructions of dynamical and stationary parameterized emission regions \citep[e.g.,][]{Broderick2016,Tiede2020}.  These are significantly simplified in the case of \VirA by the near-polar viewing angle and the approximately azimuthally symmetric, near-equatorial emission for MAD-type accretion flows \citepalias{M87_PaperV}.

Just two measurement epochs of the size of the $n=0$ and $n=1$ photon rings, associated with epochs in which the emission region has significantly varied, is sufficient to produce a joint mass and spin measurement.  This is now possible with new image analysis methods \citep[e.g.,][]{themaging}.  The accuracy of the spin measurement is strongly dependent on those of the angular radii of the $n=1$ photon ring, requiring a precision better than 1\% in practice.  Assuming that this precision can be reached in practice, a joint analysis of the 2017 and 2018 EHT observations campaign may present the first opportunity to use horizon-resolving images to measure a black hole spin.

Observations with proposed future Earth-based arrays \citep[e.g.,][]{ngEHT,ALEX} will provide much higher signal-to-noise ratios and much better baseline coverage, yielding correspondingly better estimates of the $n=0$ and $n=1$ photon ring sizes.  As a result, these future experiments hold great promise to repeatedly measure the black hole spin and mass in \VirA with high accuracy. 
Similarly, future space-based millimeter interferometers \citep[e.g.,][]{Johnson_2019} may enable the first detection of the $n=2$ photon ring, and thus a single-epoch spin measurement, and extending the science motivation for doing so.  This requires a similar effective imaging resolution as that required to unambiguously detect the $n=2$ photon ring.

This immediately raises the prospect of high-precision tests of general relativity.  A set of measurements of $\theta_{n=0}$ and $\theta_{n=1}$ over more than two epochs with differing emission locations is over-constrained.  That is, the bands of all must intersect at a single position in the $M$-$a$ plane.  This is distinct from prior constraints on general relativity in two ways.  First, it is based solely on high-resolution radio imaging data.  It does not require comparison with an ancillary mass estimate, i.e., that from the stellar dynamics observations at much larger radii.  As a result, the precision of the measurement is solely dependent on the precision of the ring size measurements.  Second, for the same reason that the spin measurement is independent of the systematic uncertainties on the mass described in \citepalias{M87_PaperVI}, the comparison is similarly devoid of these systematic uncertainties.  In this way, \VirA may generate high-precision tests of general relativity in the near future.

Additional complications, e.g., inclination, optical depth, kinematics, asymmetry, etc., may be naturally included via more complete semi-analytical model comparisons in which the gravitational lensing and source emission are modeled directly \citep[see, e.g.,][]{BroderickLoeb2006b,Broderick2016}.  Such models have already been deployed on \sgra in multiple analyses that recover spin and disk inclination \citep{Broderick2009b,Huang2009,Broderick2011,Broderick2016}.  State-of-the-art comparison frameworks include natively implemented examples of these sorts of physically-motivated parameterized emission models, e.g., that in \texttt{\sc{Themis}} \citep{THEMIS-CODE-PAPER}.  The analysis presented here is a particularly simple example of such a comparison, selecting a handful of salient features based on the simplicity of their physical interpretation.  Therefore, its success provides a reason for significant optimism regarding the prospects of these more detailed, though more expensive, modeling schemes.

\acknowledgments
We thank Michael D. Johnson and Thomas Bronzwaer for helpful comments.
This work was supported in part by Perimeter Institute for Theoretical Physics.  Research at Perimeter Institute is supported by the Government of Canada through the Department of Innovation, Science and Economic Development Canada and by the Province of Ontario through the Ministry of Economic Development, Job Creation and Trade.
A.E.B. thanks the Delaney Family for their generous financial support via the Delaney Family John A. Wheeler Chair at Perimeter Institute.
A.E.B. and P.T. receive additional financial support from the Natural Sciences and Engineering Research Council of Canada through a Discovery Grant. R.G receives additional support from the ERC synergy grant “BlackHoleCam: Imaging the Event Horizon of Black Holes” (Grant No. 610058). 
D.W.P. is supported by the NSF through grants AST-1952099, AST-1935980, AST-1828513, and AST-1440254; by the Gordon and Betty Moore Foundation through grant GBMF-5278; and in part by the Black Hole Initiative at Harvard University, which is funded by grants from the John Templeton Foundation and the Gordon and Betty Moore Foundation to Harvard University.

\bibliographystyle{aasjournal_aeb}
\bibliography{references}

\appendix

\section{Computing $\vartheta_{n}$} \label{app:vartheta}
\begin{figure*}
\begin{center}
\includegraphics[width=0.32\columnwidth]{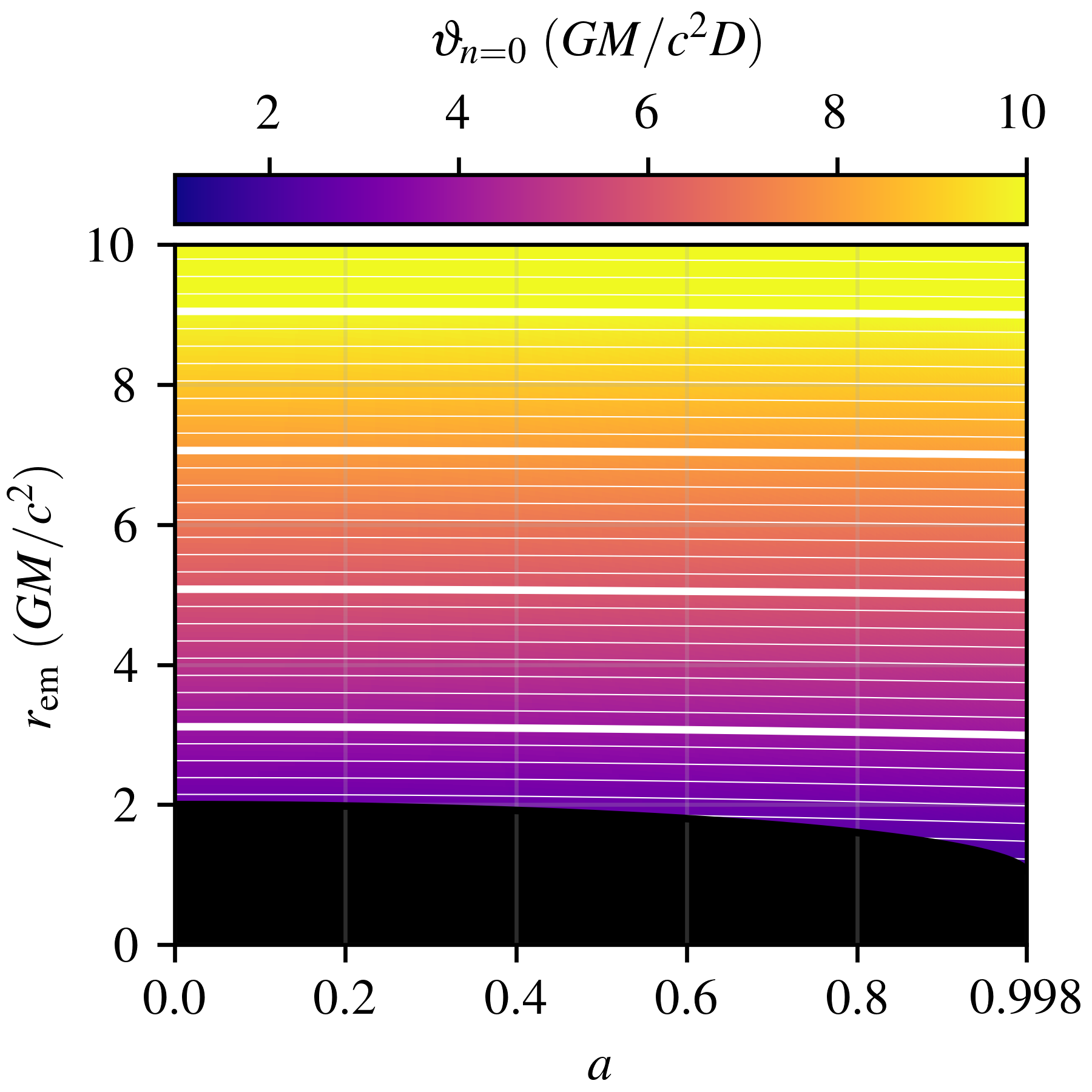}
\includegraphics[width=0.32\columnwidth]{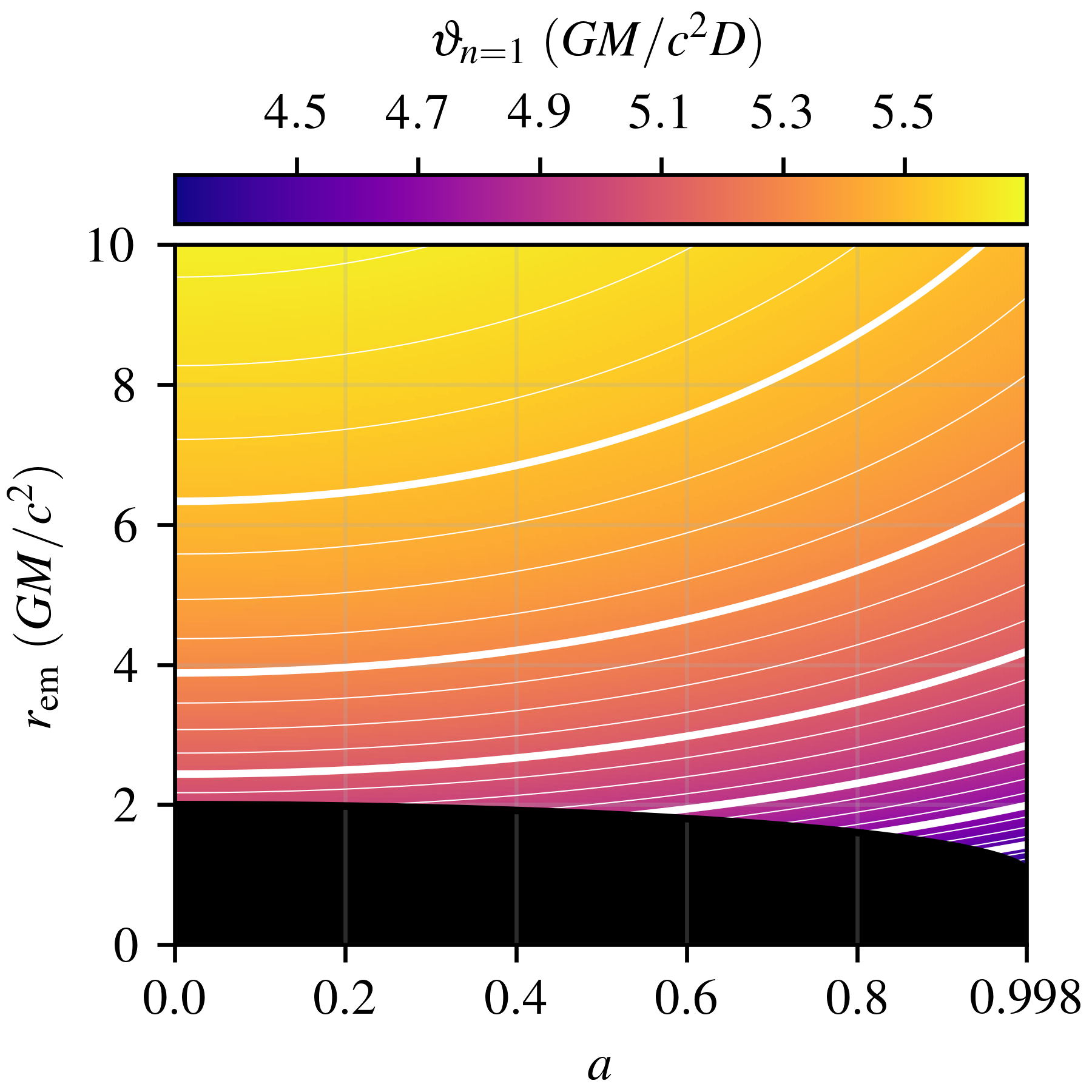}
\includegraphics[width=0.32\columnwidth]{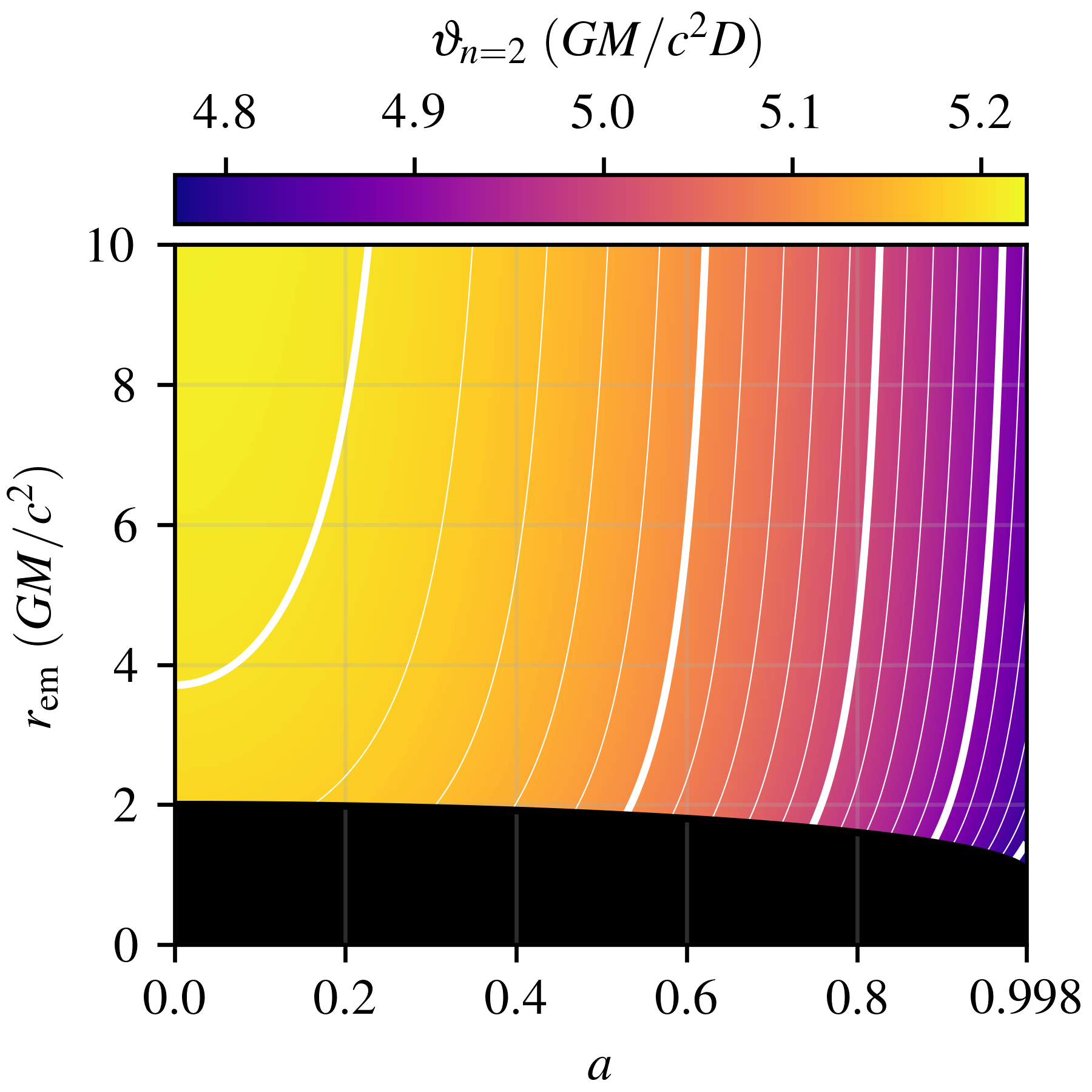}
\end{center}
\caption{$\vartheta_{n=0}$ (left), $\vartheta_{n=1}$ (center), and $\vartheta_{n=2}$ (right) as functions of $a$ and $r_{\rm em}$.  The horizon is shown by the black region.  Contours are linearly spaced and thick lines correspond to the ticks shown on the associated colorbar.}
\label{fig:theta_contours}
\end{figure*}

The relationships between photon ring radii when $n\gg1$ are derived in \citet{Johnson_2019}.  However, these asymptotic formulae are unsuitable for the low $n$ photon rings that are expected to be detected in coming years.  Therefore, we numerically construct the mapping $\vartheta_{n=0,1,2}(a,r_{\rm em})$.  These functions are shown in \autoref{fig:theta_contours}.
 
For a number of values of $a$, we begin by generating a collection of null geodesics propagated backwards in time from a set of adaptively spaced radial positions on a distant screen, located at a Boyer-Lindquist radius of $2\times10^4\,GM/c^2$ and centered on the polar axis (see upper part of \autoref{fig:geometric_bias_examples}) using the radiative transfer and ray tracing code VRT2 (\citealt{Broderick2004}, see \cite{Gold2020} for performance and comparison with alternate codes).  The resulting trajectories are inspected for crossings of the equatorial plane, at which the radial coordinate position is recorded, labeling each crossing by its order (setting $n=0$ for the last crossing, $n=1$ for the penultimate crossing, etc.).  That is, a tabulated set of $r_{{\rm em},n}(a,\vartheta)$ is generated.  Finally, this table is numerically inverted to produce a set of tabulated $\vartheta_{n}(a,r_{\rm em})$.  Repeating this procedure for many $a$ produces a set of two-dimensional tables from which values at arbitrary $a$ and $r_{\rm em}$ are obtained by interpolation.

\section{Iterative Photon Ring Peak Identification}\label{app:photon_ring_id}
Given a radial flux profile, $I(\theta)$, we seek to reconstruct the $\theta_{n=0,1,2}$.  This is modestly complicated by the overlap between photon ring fluxes.  Therefore, we implement an iterative procedure in which we identify the contribution to the flux profile of the lowest remaining order photon ring and removing it.  Here we describe this process in detail.
\begin{figure}
\begin{center}
\includegraphics[width=0.5\textwidth]{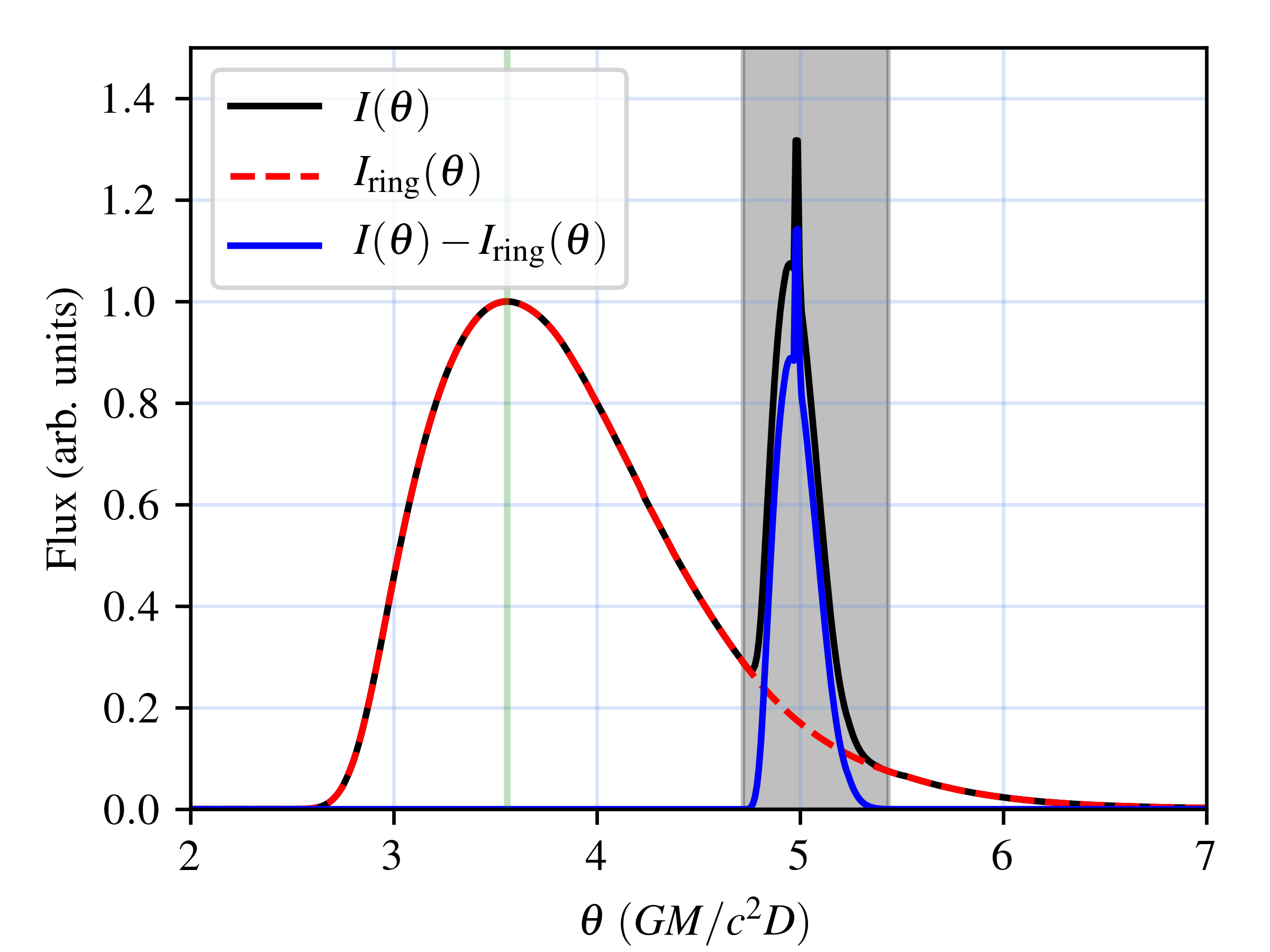}
\end{center}
\caption{Example step of the iterative peak identification and measurement procedure.  The black line shows the input radial flux distribution.  The grey band is the region between $\theta_{\rm min}$ and $\theta_{\rm max}$ that is excluded; the resulting splined $n=0$ photon ring profile is shown by the dashed red line. The angular size of the $n=0$ peak is indicated by the vertical green line.  The residual flux distribution, which is the input for the next iteration, is shown in blue.}\label{fig:iterative_rings}
\end{figure}

The procedure begins with the total $I(\theta)$ and a list of window sizes $(\theta_{\rm min},\theta_{\rm max})_{n}$ which encompass the associated order photon ring.  We then construct an approximation of the lowest order photon ring profile:
\begin{equation}
I_{{\rm ring}, n}(\theta) = 
\begin{cases}
I(\theta) & \theta<\theta_{{\rm max},n+1}~\text{and}~\theta>\theta_{{\rm min},n+1}\\
\text{cubic spline interpolation} & \text{otherwise,}
\end{cases}
\end{equation}
where the spline is performed on the values of $I(\theta)$ outside the window.  The relevant photon ring radius is set to the maximum of $I(\theta)$.  Finally, we set $I(\theta) \rightarrow I(\theta) - I_{{\rm ring},n}$, and iterate the process.  This is illustrated in \autoref{fig:iterative_rings}

The accuracy of this procedure depends on the degree of overlap and the accuracy of the cubic spline interpolations.  The latter is dependent on the widths of the photon rings (narrow interpolation regions are more accurate than broad regions) and the relative location of the rings (where rings are on top of each other, the cubic spline approximation becomes less well defined).

\end{document}